\def\aa{{A\&A}}
\def\aas{{ A\&AS}}
\def\aj{{AJ}}
\def\al{$\alpha$}
\def\bet{$\beta$}
\def\amin{$^\prime$}
\def\annrev{{ARA\&A}}
\def\apj{{ApJ}}
\def\apjs{{ApJS}}
\def\asec{$^{\prime\prime}$}

\def\deg{$^{\circ}$}

\def\e#1{$\times$10$^{#1}$}
\def\etal{{et al. }}

\def\farcs{\hbox{$.\mkern-4mu^{\prime\prime}$}}

\def\hst{{\it HST}}
\def\kms{km s$^{-1}$}
\def\lamb{$\lambda$}
\def\lax{{$\mathrel{\hbox{\rlap{\hbox{\lower4pt\hbox{$\sim$}}}\hbox{$<$}}}$}}
\def\gax{{$\mathrel{\hbox{\rlap{\hbox{\lower4pt\hbox{$\sim$}}}\hbox{$>$}}}$}}
\def\simlt{\lower.5ex\hbox{$\; \buildrel < \over \sim \;$}}
\def\simgt{\lower.5ex\hbox{$\; \buildrel > \over \sim \;$}}
\def\lum{erg s$^{-1}$}
\def\mbh{{$M_{\bullet}$}}

\def\mnras{{MNRAS}}
\def\nat{{Nature}}
\def\pasp{{PASP}}

\def\percm2{cm$^{-2}$}

\def\solmass{$M_\odot$}

\def\civ{C~{\sc iv}}
\def\mgii{Mg~{\sc ii}}
\def\feii{Fe~{\sc ii}}
\def\oi{[O~{\sc i}]}
\def\oii{[O~{\sc ii}]}
\def\oiii{[O~{\sc iii}]}

\def\heii{He~{\sc ii}}

\def\hii{H~{\sc ii}}

\def\nii{[N~{\sc ii}]}

\def\sii{[S~{\sc ii}]}

\def\lbol{$L_{{\rm bol}}$}
\def\ledd{$L_{{\rm Edd}}$}

\documentclass[cup5b]{caps}
\usepackage{graphicx}
\usepackage{ociwsymp1}
\HeadText{L. C. Ho}

\begin{document}
\pagenumbering{arabic}

\author[]{LUIS C. HO
\\The Observatories of the Carnegie Institution of Washington}

\chapter{Black Hole Demography from \\ Nearby Active Galactic Nuclei}

\begin{abstract}
A significant fraction of local galaxies show evidence of nuclear activity.  
I argue that the bulk of this activity, while energetically not remarkable, 
derives from accretion onto a central massive black hole.  The statistics 
of nearby active galactic nuclei thus provide an effective probe of 
black hole demography.  Consistent with the picture emerging from direct 
dynamical studies, the local census of nuclear activity strongly suggests that 
most, perhaps all, galaxies with a significant bulge component contain a central
massive black hole.  Although late-type galaxies appear to be generally 
deficient in nuclear black holes, there are important exceptions to this 
rule.  I highlight two examples of dwarf, late-type galaxies that 
contain active nuclei powered by intermediate-mass black holes.
\end{abstract}

\section{Introduction}

The search for massive black holes (BHs) has recently enjoyed dramatic 
progress, to the point that the statistics of BH detections have begun 
to yield useful clues on the connection between BHs and their host galaxies, 
the central theme of this Symposium.  Lest one becomes complacent, however, we 
should recognize that our knowledge of the demographics of BHs in nearby 
galaxies---on which much of the astrophysical inferences depends---remains
highly incomplete.  Direct measurements of BH masses based on resolved gas or 
stellar kinematics, while increasingly robust, are still far from routine and 
presently are available only for a limited number of galaxies (see Barth 
2004 and Kormendy 2004).  Certainly nothing approaching a ``complete'' sample 
exists yet.  More importantly, it is far from obvious that the current 
statistics are unbiased.  As discussed by Barth (2004), most nearby galaxies 
possess chaotic nuclear rotation curves that defy simple analysis.  Stellar 
kinematics provide a powerful alternative to the gas-based method, but in 
practice this technique thus far has been limited to relatively dust-free 
systems and, for practical reasons, to galaxies of relatively high central 
surface brightness.  The latter restriction selects against luminous, giant 
ellipticals.  Lastly, current surveys severely underrepresent disk-dominated 
(Sbc and later) galaxies, because the bulge component in these systems is 
inconspicuous and star formation tends to perturb the velocity field of the gas.

Given the above limitations, it would be important to consider alternative 
constraints on BH demography.  This contribution discusses the role that 
active galactic nuclei (AGNs) can play in this regard.  The commonly held, but
by now well-substantiated premise that AGNs derive their energy output from 
BH accretion implies that an AGN signifies the presence of a central BH in a 
galaxy.  The AGN signature in and of itself provides no direct information on 
BH masses, but AGN statistics can inform us, effectively and efficiently, some 
key aspects of BH demography.  For example, what fraction of all galaxies 
contain BHs?  Do BHs exist preferentially in galaxies of certain types?  Does 
environment matter?  Under what conditions do BHs light up as AGNs and how 
long does the active phase last?  What is their history of mass build-up?  
These and many other related issues are inextricably linked with the 
statistical properties of AGNs as a function of cosmological epoch.  This 
contribution concentrates on the local ($z \approx 0$) AGNs; Osmer (2004) 
considers the high-redshift population.

This review is structured as follows.  I begin with an overview of the basic 
methodology of the spectral classification of emission-line nuclei (\S 1.2) by
describing the currently adopted system, its physical motivation, the 
complications of starlight subtraction, and some practical examples.  Section 
1.3 briefly summarizes past and current spectroscopic surveys and 
introduces the Palomar survey.  The demographics of nearby AGNs is the 
subject of \S 1.4, covering detection rates based on optical surveys, 
detection rates based on radio work, the detection of weak broad 
emission lines, issues of robustness and completeless in current surveys, 
the local AGN luminosity function, the statistics of accretion luminosities,
host galaxy properties and environmental effects, and 
intermediate-mass black holes.  No discussion on nearby AGNs would be complete 
without a proper treatment on LINERs (\S 1.5).  I focus on what I 
believe are the three most important topics, namely the current evidence that 
the majority of LINERs are indeed powered by accretion, AGN photoionization 
as their dominant excitation mechanism and the demise of competing 
alternatives, and the largely still-unresolved nature of the so-called 
transition objects.  Section 1.6 gives a synopsis of the main points.

\section{Spectral Classification of Galactic Nuclei}

\subsection{Physical Motivation}

AGNs can be identified by a variety of methods.  Most AGN surveys rely on some 
aspect of the distinctive AGN spectrum, such as the presence of strong or 
broad emission lines, an unusually blue continuum, or strong radio or X-ray 
emission.  While all of these techniques are effective, none is free from 
selection effects.  To search for AGNs in nearby galaxies, where the 
nonstellar signal of the nucleus is expected to be weak relative to the host 
galaxy, the most effective and least biased method is to conduct a 
spectroscopic survey of a complete, optical-flux limited sample of galaxies.
To be sensitive to weak emission lines, the survey must be deep and of 
sufficient spectral resolution.  To obtain reliable line intensity ratios, on 
which the principal nuclear classifications are based, the data must have 
accurate relative flux calibration, and one must devise a robust scheme to 
correct for the starlight contamination. These issues are discussed below. But 
first, I must cover some basic material on spectral classification.

The most widely used system of spectral classification of emission-line nuclei 
follows the method outlined by Baldwin, Phillips, \& Terlevich (1981), and 
later modified by Veilleux \& Osterbrock (1987).  The basic idea is that the 
relative strengths of certain prominent emission lines can be used to probe
the nebular conditions of a source.  In the context of the present discussion, 
the most important diagnostic is the source of excitation, which broadly falls
into two categories: stellar photoionization or photoionization by a centrally 
located, spectrally hard radiation field, such as that produced by the 
accretion disk of a massive BH.  The latter class of sources are generically 
called AGNs, which are most relevant to issues of BH demography.

How does one distinguish stellar from nonstellar photoionization?  The 
forbidden lines of the doublet \oi\ \lamb\lamb 6300, 6364 rise from 
collisional excitation of O$^0$ by hot electrons.  Since the ionization 
potential of O$^0$ (13.6 eV) is nearly identical to that of hydrogen, in an 
ionization-bounded nebula \oi\ is produced predominantly in the ``partially 
ionized zone,'' wherein both neutral oxygen and free electrons coexist.  In 
addition to O$^0$, the conditions of the partially ionized zone are also 
favorable for S$^+$ and N$^+$, whose ionization potentials are 23.3 eV and 
29.6 eV, respectively.  Hence, in the absence of abundance anomalies, \nii\ 
\lamb\lamb 6548, 6583 and \sii\ \lamb\lamb 6716, 6731 are strong (relative to, 
say, H\al) whenever \oi\ is strong, and {\it vice versa}.  

In a nebula photoionized by young, massive stars, the partially ionized zone 
is very thin because the ionizing spectrum of OB stars contains few photons 
with energies greater than 1 Rydberg.  Hence, in the optical spectra of \hii\ 
regions and starburst nuclei (hereinafter \hii\ nuclei\footnote{As originally 
defined (Weedman et al. 1981; Balzano 1983), a star{\it burst}\ nucleus is one 
whose current star formation rate is much higher than its past average rate. 
This terminology presupposes knowledge of the star formation history of the 
system. Since this information is usually not available for any individual 
object, I will adopt the more general designation of ``\hii\ nucleus.''}) 
the low-ionization transitions \nii, \sii, and especially \oi\ are very weak.  
By contrast, a harder radiation field, such as that of an AGN power-law 
continuum that extends into the extreme-ultraviolet (UV) and X-rays, penetrates 
much deeper into an optically thick cloud.  X-ray photoionization and 
Auger processes release copious hot electrons in this predominantly neutral
region, creating an extensive partially ionized zone.  The spectra of AGNs, 
therefore, exhibit relatively strong low-ionization forbidden lines.  

\subsection{Sample Spectra}
The spectra shown in Figure 1.1 illustrate the empirical distinction between 
AGNs and \hii\ nuclei.  In NGC 7741, which has a well-known starburst nucleus 
(Weedman et al. 1981), \oi, \nii, and \sii\ are weak relative to H\al.  The 
\oiii\ \lamb\lamb 4959, 5007 doublet is quite strong compared to \oii\ \lamb 
3727 or H\bet\ because the metal abundance of NGC 7741's nucleus is rather 
low, although the ionization level of \hii\ nuclei can span a wide range, 
depending on metallicity (Ho, Filippenko, \& Sargent 1997c).  On the other 
hand, the low-ionization lines are markedly stronger in the other two 
objects shown, both of which qualify as AGNs.  NGC 1358 is an example of a 
galaxy with a ``high-ionization'' AGN or ``Seyfert'' nucleus.  NGC 1052 is the 
prototype of the class known as ``low-ionization nuclear emission-line 
regions'' or ``LINERs.''  The ionization level can be judged by the relative 
strengths of the oxygen lines, but in practice is most easily gauged by the
\oiii/H\bet\ ratio.  In the commonly adopted system of Veilleux \& Osterbrock 
(1987), the division between Seyferts and LINERs occurs at \oiii\ \lamb 
5007/H\bet\ = 3.0.  Ho, Filippenko, \& Sargent (2003) stress, however, that 
this boundary has no strict physical significance.  The ionization level of the 
narrow-line region (NLR) in large, homogeneous samples of AGNs spans a wide and 
apparently continuous range; contrary to the claims of some studies (e.g., 
V\'eron-Cetty \& V\'eron 2000) there is no evidence for any clear-cut 
transition between Seyferts and LINERs (Ho et al. 2003; Heckman 2004).

\begin{figure*}[t]
\includegraphics[width=0.77\columnwidth,angle=-90,clip]{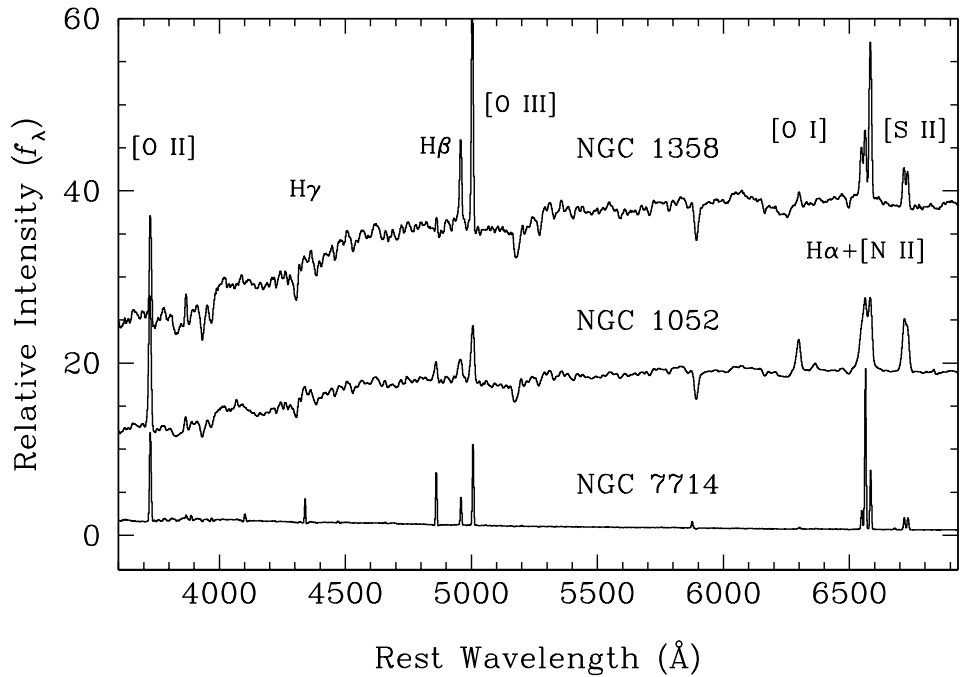}
\vskip 0pt \caption{
Sample optical spectra of the various classes of emission-line nuclei.  NGC 
1358 = Seyfert; NGC 1052 = LINER; NGC 7714 = \hii.   The prominent emission 
lines are identified.  (Based on Ho et al. 1993a and unpublished data.) 
\label{fig1}}
\end{figure*}

\begin{figure*}[t]
\includegraphics[width=0.77\columnwidth,angle=-90,clip]{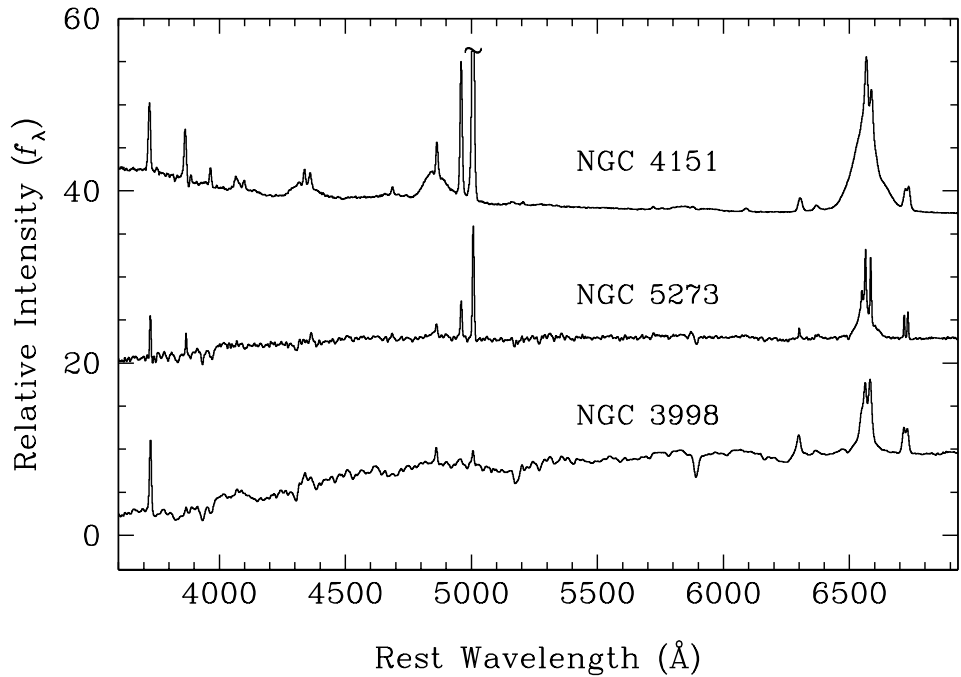}
\vskip 0pt \caption{
Sample optical spectra of broad-line AGNs.  NGC 4151 = ``classical'' Seyfert~1;
NGC 5273 = typical low-luminosity Seyfert~1; NGC 3998 = LINER~1.   
(Based on Ho et al. 1993a and unpublished data.)
\label{fig2}}
\end{figure*}

The classification system discussed above makes no reference to the profiles
of the emission lines.  Luminous AGNs such as quasars and many ``classical''
Seyfert galaxies exhibit permitted lines with a characteristically broad
component, with FWHM widths of $\sim 1000-10,000$ \kms.  This component arises
from the broad-line region (BLR), which is thought to be physically
distinct from the NLR responsible for the narrow lines.  Following Khachikian
\& Weedman (1974), it is customary to refer to Seyferts with and without
(directly) detectable broad lines as ``type~1'' and ``type~2'' sources,
respectively.   As discussed in \S~1.4.3, this nomenclature can also be 
extended to include LINERs, which also contain broad emission lines. 
Figure 1.2 gives some examples.  The spectrum of the bright Seyfert galaxy 
NGC 4151 is familiar to all: strong, broad permitted lines superposed on an 
unambiguous featureless, nonstellar blue continuuum.  But this object is not 
typical.  Even within the Seyfert class, most objects resemble NGC 5273, where 
the broad component is easily visible only for H\al\ and the featureless 
continuum is heavily diluted by the host galaxy light.  The same applies to 
LINERs (e.g., NGC 3998), where the host galaxy dilution is even more extreme; 
nonetheless, with careful starlight subtraction (\S~1.2.4) and profile 
modeling (\S~1.4.3), one can detect broad H\al\ emission in many LINERs.

\subsection{Diagnostic Diagrams}

The classification system of Veilleux \& Osterbrock (1987), which I 
adopt throughout this paper, is based on two-dimensional line-intensity ratios
constructed from \oiii\ \lamb 5007, H\bet\ \lamb 4861, \oi\ \lamb 6300, H\al\ 
\lamb6563, \nii\ \lamb 6583, and \sii\ \lamb\lamb 6716, 6731 (here H\bet\ and 
H\al\ refer only to the narrow component of the line).  The main virtues of 
this system, shown in Figure 1.3, are (1) that it uses relatively strong lines, 
(2) that the lines lie in an easily accessible region of the optical spectrum, 
and (3) that the line ratios are relatively insensitive to reddening 
corrections because of the close separation of the lines.  The 
definitions of the various classes of emission-line objects 
are given in Ho, Filippenko, \& Sargent (1997a)\footnote{The classification 
criteria adopted here differ slightly, but not appreciably, from those 
proposed by Kewley et al. (2001) based on theoretical models.}.  In 
addition to the three main classes discussed thus far---\hii\ nuclei, 
Seyferts, and LINERs---Ho, Filippenko, \& Sargent (1993a) identified a group 
of ``transition objects'' whose \oi\ strengths are intermediate between those 
of \hii\ nuclei and LINERs.  Since they tend to emit weaker \oi\ emission than 
classical LINERs, previous authors have called them ``weak-\oi\ LINERs'' 
(Filippenko \& Terlevich 1992; Shields 1992; Ho \& Filippenko 1993).  Ho et al. 
(1993a) postulated that transition objects are composite systems having both an 
\hii\ region and a LINER component; I will return to the nature of these 
sources in \S~1.5.3.

I note that my definition of LINERs differs from that originally proposed by 
Heckman (1980b), who used solely the oxygen lines: \oii\ \lamb 3727 $>$ 
\oiii\ \lamb 5007 and \oi\ \lamb 6300 $>$ 0.33 \oiii\ \lamb 5007.  The two 
definitions, however, are nearly equivalent.  Inspection of the full optical 
spectra of Ho et al. (1993a), for example, reveals that emission-line nuclei 
classified as LINERs based on the Veilleux \& Osterbrock diagrams almost 
invariably also satisfy Heckman's criteria.  This is a consequence of the 
inverse correlation between \oiii/H\bet\ and \oii/\oiii\ in photoionized gas 
with fairly low excitation (\oiii/H\bet\ \lax 3; see Fig. 2 in Baldwin et al. 
1981).

\begin{figure*}[t]
\includegraphics[width=1.00\columnwidth,angle=0,clip]{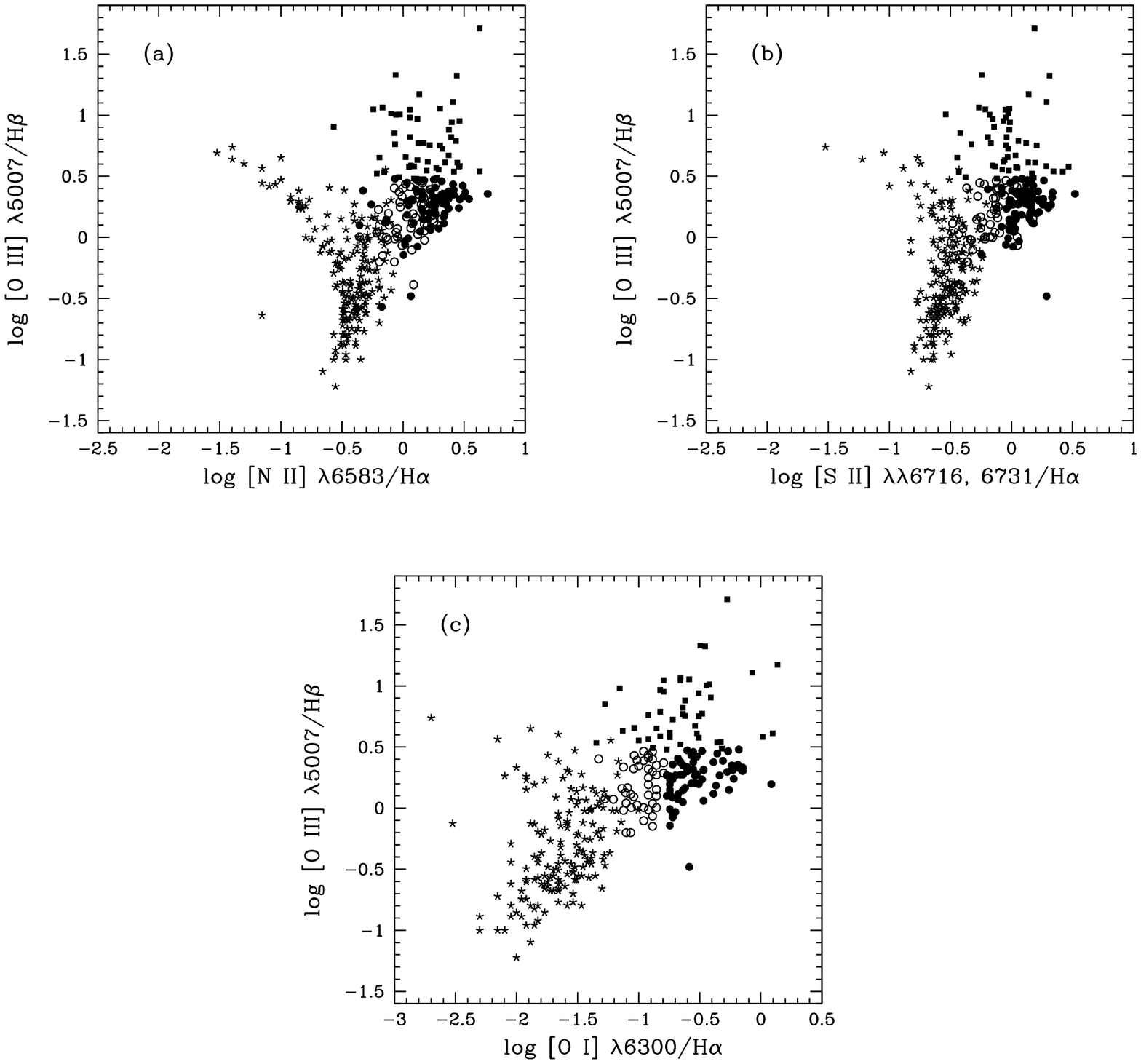}
\vskip 0pt \caption{
Diagnostic diagrams plotting ({\it a}) log \oiii\ \lamb 5007/H\bet\
versus log \nii\ \lamb 6583/H\al, ({\it b}) log \oiii\ \lamb 5007/H\bet\
versus log \sii\ \lamb\lamb 6716, 6731/H\al, and ({\it c}) log \oiii\ \lamb
5007/H\bet\ versus log \oi\ \lamb 6300/H\al.  The nuclear spectral classes
shown are \hii\ nuclei ({\it asterisks}), Seyfert nuclei ({\it squares}), LINERs
({\it solid circles}), and transition objects ({\it open circles}).  (Adapted 
from Ho et al. 1997a.)
\label{fig3}}
\end{figure*}

\subsection{Starlight Subtraction}

The scheme outlined above, while conceptionally simple, overlooks one key 
practical complication.  The integrated spectra of galactic nuclei include 
emission from stars, which in most nearby systems overwhelms the nebular line 
emission.  This can be seen in Figure 1.1, or from a cursory examination of the 
spectral atlas of Ho, Filippenko, \& Sargent (1995).  Any reliable measurement 
of the emission-line spectrum of galactic nuclei, therefore, {\it must}\ 
properly account for the starlight contamination.

An effective strategy for removing the starlight from an integrated spectrum
is that of ``template subtraction,'' whereby a template spectrum devoid of
emission lines is suitably scaled to and subtracted from the spectrum of
interest to yield a continuum-subtracted, pure emission-line spectrum.  A
number of approaches have been adopted to construct the template.  These 
include (1) using the spectrum of an off-nuclear position within the same 
galaxy (e.g., Storchi-Bergmann, Baldwin, \& Wilson 1993); (2) using the 
spectrum of a different galaxy devoid of emission lines (e.g., Costero \& 
Osterbrock 1977; Filippenko \& Halpern 1984; Ho et al. 1993a); (3) using a 
weighted linear combination of the spectra of a number different galaxies, 
chosen to best match the stellar population and velocity dispersion (Ho et al. 
1997a); (4) using the spectrum derived from a principal-component analysis of 
a large set of galaxies (Hao \& Strauss 2004); and (5) using a model spectrum 
constructed from population synthesis techniques, using as input a library of 
spectra of either individual stars (e.g., Keel 1983a) or star clusters 
(e.g., Bonatto, Bica, \& Alloin 1989; Raimann et al. 2001).

Figure 1.4 illustrates the starlight subtraction process for the \hii\ nucleus 
in NGC 3596 and for the Seyfert~2 nucleus in NGC 7743, using the method of Ho 
et al. (1997a).  Given a list of input spectra derived from galaxies devoid of 
emission lines and an initial guess of the velocity dispersion, a 
$\chi^2$-minimization algorithm solves for the systemic velocity, the 
line-broadening function, the relative contribution of the various input 
spectra, and the general continuum shape.  The best-fitting model is then 
subtracted from the original spectrum, yielding a pure emission-line spectrum.  
In the case of NGC 3596, the model consisted of the combination of the spectrum
of NGC 205, a dE5 galaxy with a substantial population of A-type stars, 
and NGC 4339, an E0 galaxy having a K-giant spectrum. Note that in the original 
observed spectrum (top), H$\gamma$, \oiii\ \lamb\lamb 4959, 5007, and \oi\ 
\lamb 6300 were hardly visible, whereas after starlight subtraction (bottom) 
they can be easily measured.  The intensities of both H\bet\ and H\al\ have 
been modified substantially, and the ratio of the two \sii\ \lamb\lamb 6716, 
6731 lines changed.  The effective template for NGC 7743 made use of NGC 205, 
NGC 4339, and NGC 628, an Sc galaxy with a nucleus dominated by A and F stars.

Some studies (e.g., Kim et al. 1995) implicitly assume that only the hydrogen 
Balmer lines are contaminated by starlight, and that the absorption-line 
component can be removed by subtracting a constant equivalent width 
(2--3 \AA).  This procedure is inadequate for a number of reasons.  First, the 
stellar population of nearby galactic nuclei, although relatively uniform, is 
by no means invariant (Ho et al. 2003).  Second, the equivalent widths of the 
different Balmer absorption lines within each galaxy are generally not 
constant.  Third, the Balmer absorption lines affect not only the strength but 
also the shape of the Balmer emission lines.  And finally, as the above 
examples show, starlight contaminates lines other than just the Balmer lines.

\begin{figure*}[t]
\includegraphics[width=1.00\columnwidth,angle=0,clip]{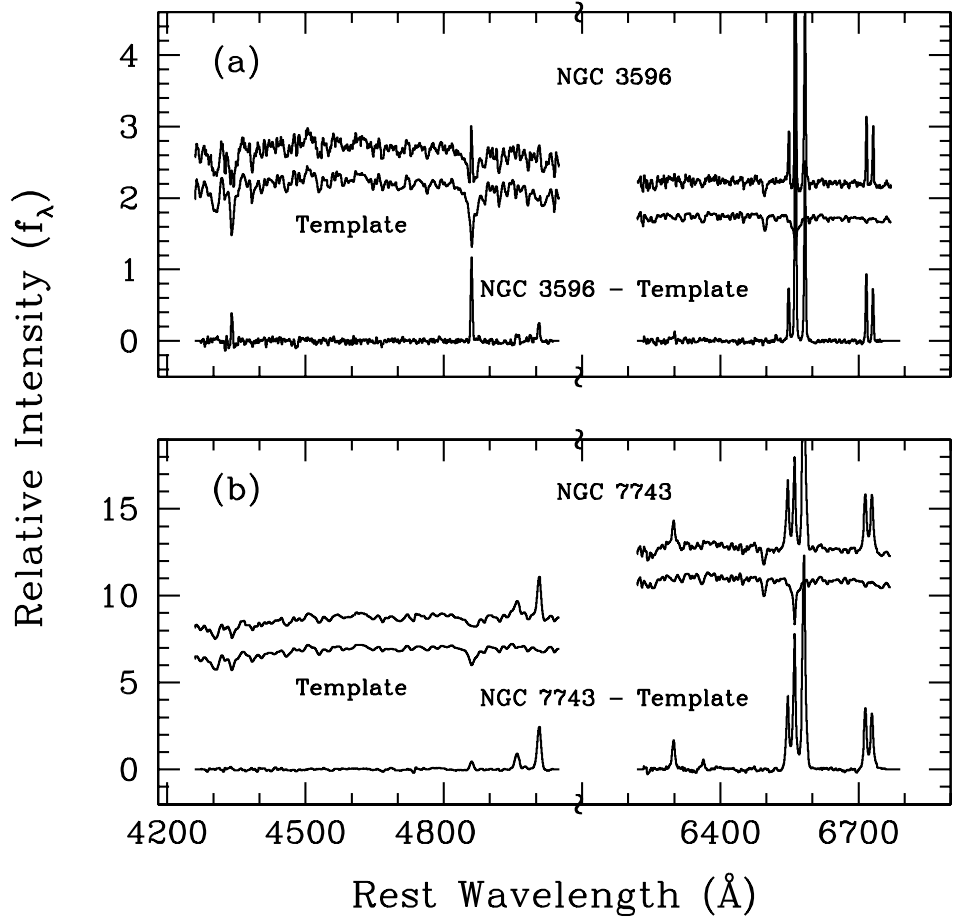}
\vskip 0pt \caption{
Illustration of the method of starlight subtraction.  In each panel, 
the top plot shows the observed spectrum, the middle plot the best-fitting 
``template'' used to match the stellar component, and the bottom plot the 
difference between the object spectrum and the template.  In the case of 
NGC 3596 ({\it a}), the model was constructed from NGC 205 and NGC 4339, while 
for NGC 7743 ({\it b}), the model was derived from a linear combination of 
NGC 205, NGC 4339, and NGC 628.  (Adapted from Ho et al. 1997a.) 
\label{fig4}}
\end{figure*}

\section{Spectroscopic Surveys of Nearby Galactic Nuclei}

It was apparent from some of the earliest redshift surveys that the 
central regions of galaxies often show evidence of strong emission lines 
(e.g., Humason, Mayall, \& Sandage 1956).  A number of studies also indicated 
that in many instances the spectra revealed abnormal line-intensity ratios, 
most notably the unusually great strength of \nii\ relative to H\al\ 
(Burbidge \& Burbidge 1962, 1965; Rubin \& Ford 1971).  That the optical 
emission-line spectra of some nuclei show patterns of low ionization was 
noticed from time to time, primarily by Osterbrock and his colleagues (e.g., 
Osterbrock \& Miller 1975; Koski \& Osterbrock 1976; Costero \& Osterbrock 
1977; Grandi \& Osterbrock 1978; Phillips 1979), but also by others (e.g., 
Disney \& Cromwell 1971; Danziger, Fosbury, \& Penston 1977; Fosbury \etal 
1977, 1978; Penston \& Fosbury 1978; Stauffer \& Spinrad 1979).

Most of the activity in this field culminated in the 1980s, beginning with the 
recognition (Heckman, Balick, \& Crane 1980; Heckman 1980b) of LINERs as a 
major constituent of the extragalactic population, and then followed by further 
systematic studies of larger samples of galaxies (Stauffer 1982a, b; Keel 
1983a, b; Phillips \etal 1986; V\'eron \& V\'eron-Cetty 1986; V\'eron-Cetty \& 
V\'eron 1986; see Ho 1996 for more details).  These surveys established three 
important results: (1) a large fraction of local galaxies contain 
emission-line nuclei; (2) many of these sources are LINERs; and (3) LINERs may 
be accretion-powered systems.

Despite the success of these seminal studies, there was room for improvement.
Although most of the surveys attempted some form of starlight subtraction, the 
accuracy of the methods used tended to be fairly limited (see discussion in 
Ho et al. 1997a), the procedure was sometimes inconsistently applied, and in 
some of the surveys starlight subtraction was largely neglected.  The problem 
is exacerbated by the fact that the apertures used for the observations 
were quite large, thereby admitting an unnecessarily large amount of 
starlight.  Furthermore, most of the data were collected with rather poor 
spectral resolution (FWHM $\approx$ 10 \AA).  Besides losing useful 
kinematic information, blending between the emission and absorption 
components further compromises the ability to separate the two.

Thus, it is clear that much would be gained from a survey having greater 
sensitivity to the detection of emission lines.  The sensitivity can be 
improved in at least four ways---by taking spectra with higher 
signal-to-noise ratio and spectral resolution, by using a narrower slit 
to better isolate the nucleus, and by employing more effective methods to 
handle the starlight correction.  

The Palomar spectroscopic survey of nearby galaxies (Filippenko \& Sargent 
1985, 1986; Ho et al. 1995, 1997a--e, 2003) was designed with these goals in 
mind.  Using a double CCD spectrograph mounted on the Hale 5-m reflector at 
Palomar Observatory, high-quality, moderate-resolution, long-slit spectra were 
obtained for a magnitude-limited ($B_T\,\leq$ 12.5 mag) sample of 486 northern 
($\delta$ $>$ 0\deg) galaxies.  The spectra simultaneously cover the wavelength
ranges 6210--6860 \AA\ with $\sim$2.5 \AA\ resolution (FWHM) and 4230--5110 
\AA\ with $\sim$4 \AA\ resolution.  Most of the observations were obtained 
with a narrow slit (generally 2\asec, and occasionally 1\asec), and the 
exposure times were suitably long (up to 1 hr or more for some objects with 
low central surface brightness) to secure data of high signal-to-noise ratio.  
This survey contains the largest database to date of homogeneous and 
high-quality optical spectra of nearby galaxies.  It is also the most 
sensitive; the detection limit for emission lines is $\sim$0.25 
\AA, roughly an order-of-magnitude improvement compared to previous work.  The 
selection criteria of the survey ensure that the sample gives a fair 
representation of the local ($z\,\approx\,0$) galaxy population, and the 
proximity of the objects (median distance = 17 Mpc) enables relatively good 
spatial resolution to be achieved (typically \lax 200 pc).  These properties 
of the Palomar survey make it ideally suited to address issues on the 
demographics and physical properties of nearby, and especially low-luminosity, 
AGNs.  Unless otherwise noted, the main results presented in the rest of this 
paper will be taken from the Palomar survey.

\section{Demographics of Nearby AGNs}

\subsection{Detection Rates}

\begin{figure*}[t]
\includegraphics[width=0.515\columnwidth,angle=270,clip]{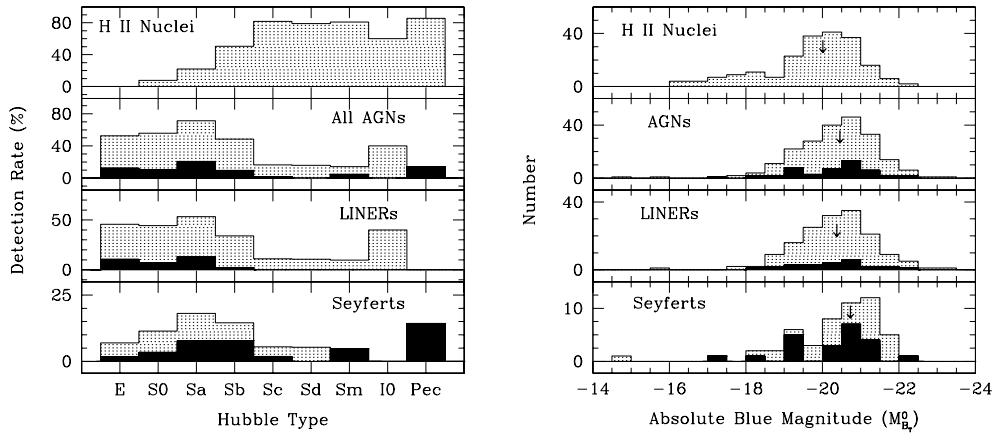}
\vskip 0pt \caption{
({\it Left}) Percentage of galaxies with the various classes of
emission-line nuclei detected as a function of Hubble type. ({\it Right})
Distribution of the classes of emission-line nuclei as a function of the
absolute $B$ magnitude of the host galaxy. The arrow in each panel marks
the median of the distribution.  The solid and hatched histograms denote
type~1 and type~1 + type~2 sources, respectively.
(Adapted from Ho et al.  1997b.) 
\label{fig5}}
\end{figure*}

In qualitative agreement with previous surveys, the Palomar survey finds that 
a substantial fraction (86\%) of all galaxies contain detectable emission-line 
nuclei (Ho et al. 1997b).  The detection rate is essentially 100\% for all 
disk (S0 and spiral) galaxies, and over 50\% for elliptical galaxies.  One of 
the most surprising results is the large fraction of objects classified as 
AGNs or AGN candidates.  Summed over all Hubble types, 43\% of all galaxies 
that fall in the survey limits can be considered ``active'' (Fig. 1.5).  
This percentage becomes even more remarkable for galaxies with an obvious bulge 
component, rising to $\sim$50\%--70\% for Hubble types E--Sbc.  By contrast, 
the detection rate of AGNs drops dramatically toward later Hubble types (Sc 
and later), which almost invariably (80\%) host \hii\ nuclei.  This strong 
dependence of nuclear spectral class on Hubble type has been noticed in 
earlier studies (Heckman 1980a; Keel 1983a; Terlevich, Melnick, \& Moles 1987).

Among the active sources, 11\% have Seyfert nuclei, at least doubling older 
estimates (Stauffer 1982b; Keel 1983b; Phillips, Charles, \& Baldwin 1983; 
Huchra \& Burg 1992).  LINERs constitute the dominant population of AGNs.  
``Pure'' LINERs are present in $\sim$20\% of all galaxies, whereas transition 
objects, which by assumption also contain a LINER component, account for 
another $\sim$13\%. Thus, if all LINERs can be regarded as genuine AGNs (see 
\S~1.5), they truly are the most populous constituents---they make up 1/3 of 
all galaxies and 2/3 of the AGN population (here taken to mean all objects 
classified as Seyferts, LINERs, and transition objects).

The sample of nearby AGNs emerging from the Sloan Digital Sky Survey (SDSS)
(Kauffmann et al. 2003; Hao \& Strauss 2004; Heckman 2004) far surpasses that 
of the Palomar survey in number.  Within the magnitude range $14.5 < r < 17.7$, 
Kauffmann et al. (2003) report an overall AGN fraction (for narrow-line 
sources) of $\sim$40\%, of which $\sim$10\% are Seyferts, the rest LINERs 
and transition objects.  Using a different method of starlight subtraction, 
Hao \& Strauss (2004) obtain very similar statistics for their sample of 
Seyfert galaxies.  Although these detection rates broadly resemble those 
of the Palomar survey, one should recognize important differences between the 
two surveys.  The Palomar objects extend much farther down the luminosity 
function than the SDSS.  The emission-line detection limit of the Palomar 
survey, 0.25 \AA, is roughly 10 times fainter than the cutoff 
chosen by Hao \& Strauss (2004). The faint end of the Palomar H\al\ luminosity 
function reaches 1\e{38} \lum, again a factor of 10 lower than the SDSS
counterpart.  Moreover, the 3\asec-diameter fibers used in the SDSS 
subtend a physical scale of $\sim$5.5 kpc at the typical redshift
$z \approx 0.1$, 30 times larger than in the Palomar survey.  The SDSS
spectra, therefore, include substantial contamination from off-nuclear 
emission, which would dilute, and in some cases inevitably confuse, the signal 
from the nucleus.  

Contamination by host galaxy emission has two consequences.  First, only 
relatively bright nuclei have enough contrast to be detected; this is 
consistent with the sensitivity difference described above.  But second, it 
can introduce a more pernicious systematic effect that can be hard to 
quantify.  Apart from normal \hii\ regions, galactic disks are known to 
contain emission-line regions that exhibit low-ionization, LINER-like spectra, 
which can be confused with genuine {\it nuclear}\ LINERs.  Examples include 
gas shocked by supernova remnants (e.g., Dopita \& Sutherland 1995), ejecta 
from starburst-driven winds (Armus, Heckman, \& Miley 1990), and diffuse 
ionized plasma (e.g., Lehnert \& Heckman 1994; Collins \& Rand 2001).  
Massive, early-type galaxies, though generally lacking in ongoing star 
formation, do often possess X-ray emitting atmospheres that exhibit extended, 
low-ionization emission-line nebulae (e.g., Fabian et al. 1986; Heckman et al. 
1989).  These physical processes, while interesting in their own right, are 
not directly related, and thus irrelevant, to the AGN phenomenon.  Thus, 
``LINERs'' selected from samples of distant galaxies should be regarded with 
considerable caution.

\subsection{Statistics from Radio Surveys}

The prevalence of weak AGNs in nearby galaxies is corroborated by 
high-resolution radio continuum surveys.  Sadler, Jenkins, \& Kotanyi (1989) 
and Wrobel \& Heeschen (1991) report a relatively high incidence ($\sim$50\%) 
of compact radio cores in complete, optical flux-limited samples of elliptical 
and S0 galaxies.  The radio powers are quite modest, generally in the range of 
$10^{19}-10^{21}$ W Hz$^{-1}$ at 5 GHz.  When available, the spectral indices 
tend to be relatively flat (e.g., Slee et al. 1994).  The optical counterparts 
of the radio cores are usually spectroscopically classified as LINERs 
(Phillips et al. 1986; Ho 1999a). 
 
\subsection{Broad Emission Lines}

\begin{figure*}[t]
\includegraphics[width=0.49\columnwidth,angle=0,clip]{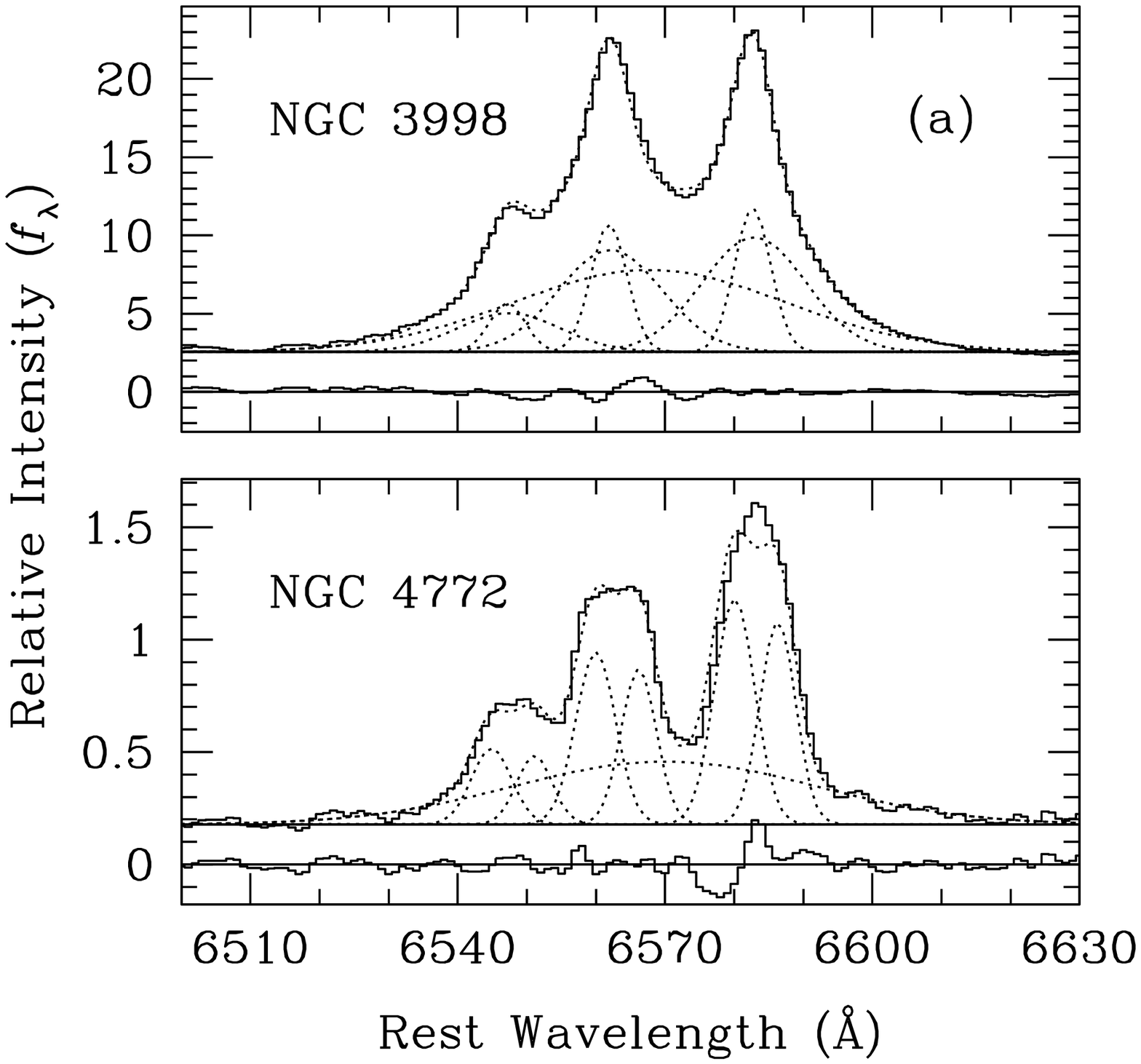}
\includegraphics[width=0.49\columnwidth,angle=0,clip]{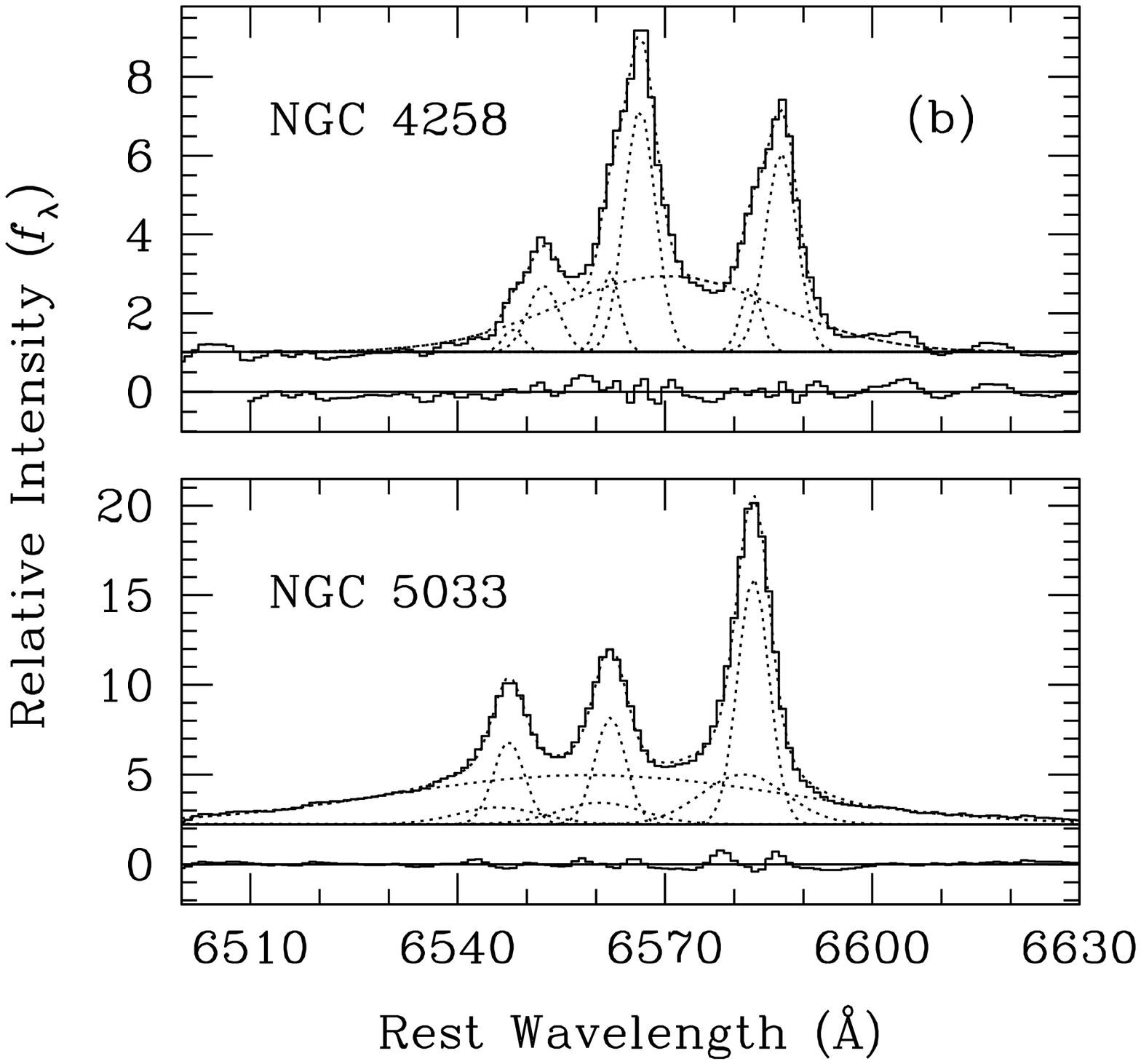}
\vskip 0pt \caption{
Examples of ({\it a}) LINERs and ({\it b}) Seyferts with broad H\al\
emission.  \nii\ \lamb\lamb6548, 6583 and the narrow component of H\al\ are
assumed to have the same shape as \sii\ \lamb\lamb6716, 6731, and the broad
component of H\al\ is modeled as a single Gaussian.  Residuals of the fit are
shown on the bottom of each panel.  (Adapted from Ho et al. 1997e.)
\label{fig6}}
\end{figure*}

Broad emission lines, a defining attribute of classical Seyferts and quasars, 
are also found in nuclei of much lower luminosities.  The well-known case of 
the nucleus 
of M81 (Peimbert \& Torres-Peimbert 1981; Filippenko \& Sargent 1988), for 
example, has a broad (FWHM $\approx$ 3000 \kms) H\al\ line with a luminosity of 
only $2 \times\,10^{39}$ \lum\ (Ho, Filippenko, \& Sargent 1996), and many 
other less conspicuous cases have been discovered in the Palomar survey (Ho et 
al. 1997e; Fig. 1.6).  Searching for broad H\al\ emission in nearby nuclei is 
nontrivial, because it entails measurement of a (generally) weak, low-contrast,
broad emission feature superposed on a complicated stellar background.  Thus,
the importance of careful starlight subtraction cannot be overemphasized.
Moreover, even if one were able to perfectly remove the starlight, one still
has to contend with deblending the H\al\ + \nii\ \lamb\lamb6548, 6583
complex.  The narrow lines in this complex are often heavily blended together,
and rarely do the lines have simple profiles.  The strategy adopted by Ho 
\etal (1997e) is to use the empirical line profile of the \sii\ lines to model 
\nii\ and the narrow component of H\al.

Of the 221 emission-line nuclei in the Palomar survey classified as LINERs, 
transition objects, and Seyferts, 33 (15\%) definitely have broad H\al, and an
additional 16 (7\%) probably do.  Questionable detections were found in
another 8 objects (4\%).  Thus, approximately 20\%--25\% of all nearby AGNs 
are type~1 sources.  These numbers, of course, should be regarded as lower 
limits, since undoubtedly there must exist AGNs with even weaker broad-line 
emission that fall below the detection threshold.

It is illuminating to consider the incidence of broad H\al\ emission as
a function of spectral class.  Among objects formally classified as Seyferts 
(according to their narrow-line spectrum), approximately 40\% are Seyfert~1s.  
The implied ratio of Seyfert~1s to Seyfert~2s (1:1.6) has important 
consequences for some models concerning the evolution and small-scale geometry 
of AGNs (e.g., Osterbrock \& Shaw 1988; Lawrence 1991).  Despite claims to the 
contrary (Krolik 1998; Sulentic, Marziani, \& Dultzin-Hacyan 2000), broad 
emission lines emphatically are {\it not}\ exclusively confined to Seyfert 
nuclei.  Within the Palomar sample, nearly 25\% of the ``pure'' LINERs have 
detectable broad H\al\ emission.  By direct analogy with the familiar 
nomenclature established for Seyferts, LINERs can be divided 
into ``type~1'' and ``type~2'' sources according to the presence or absence of
broad-line emission, respectively (Ho et al. 1997a, 1997e).  The detection 
rate of broad H\al, however, drops drastically for transition objects.  The 
cause for this dramatic change is unclear, but a possible explanation is that 
the broad-line component is simply too weak to be detected in the presence of 
substantial contamination from the \hii\ region component.  

A subset of LINERs contain broad lines with {\it double-peaked}\ profiles 
(Fig. 1.7), analogous to those seen in a minority of radio galaxies (Eracleous 
\& Halpern 1994), where they are often interpreted as a kinematic signature of 
a relativistically broadened accretion disk (Chen \& Halpern 1989).  Most 
of the nearby cases have been discovered serendipitously, either as a result 
of the broad component being variable (e.g., Storchi-Bergmann et al. 1993) or 
because of the increased sensitivity to weak, broad features afforded by 
small-aperture measurements made with the {\it Hubble Space Telescope (HST)}\ 
(Shields et al. 2000; Ho et al. 2000, and references therein).

\begin{figure*}[t]
\includegraphics[width=0.72\columnwidth,angle=-90,clip]{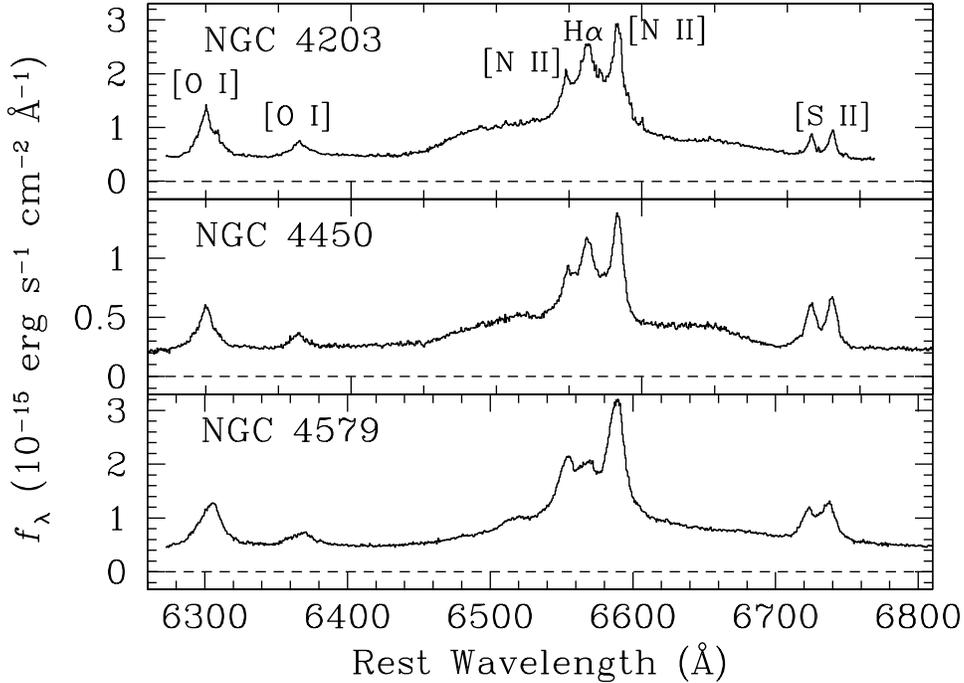}
\vskip 0pt \caption{
Examples of LINERs with broad, double-peaked H\al\ emission discovered
with {\it HST}.  (Adapted from Ho et al. 2000, Shields et al. 2000, and 
Barth et al. 2001.)
\label{fig7}}
\end{figure*}

\subsection{Robustness and Completeness}

To gain confidence in the current AGN statistics, one must have some handle on 
whether the existing AGN detections are trustworthy and whether there are 
many AGNs that have been missed.  The robustness issue hinges on the question 
of whether the weak, nearby sources classified as AGNs are truly 
accretion-powered.  As summarized in \S~1.5, this appears to be largely the 
case.  The completeness issue can be examined in two regimes.  Among bulged 
(Sbc and earlier) galaxies, for which the spectroscopic AGN fractions are 
already very high ($\sim$50\%--75\%; Fig. 1.5), there is no room for a large
fraction of missing AGNs.  The same does not necessarily hold for galaxies of 
Hubble types Sc and later.  While the majority of these systems are 
spectroscopically classified as \hii\ nuclei, one must be wary that weak AGNs, 
if present, may be masked by brighter off-nuclear \hii\ regions or \hii\ 
regions projected along the line of sight.  After all, some very late-type 
galaxies {\it do}\ host {\it bona fide}\ AGNs (see \S 1.4.8).

The AGN content of late-type galaxies can be independently assessed by using 
a diagnostic less prone to confusion by star-forming regions.  The presence of 
a compact, nuclear radio or X-ray core turns out to be a useful AGN filter, 
since genuine AGNs almost always possess compact emission in these bands.  
Because of the expected weakness of the nuclei, however, any search for core 
emission must be conducted at relatively high sensitivity and angular 
resolution (\lax 1\asec).  In practice, this requires {\it Chandra} for the 
X-rays and an interferometer such as the Very Large Array (VLA) for the radio.

Ulvestad \& Ho (2002) have performed a VLA survey for radio cores in a 
distance-limited sample of 40 Palomar Sc galaxies classified as hosting \hii\ 
nuclei.  To a sensitivity limit of $P_{\rm 6 cm} \approx 10^{18}-10^{20}$ 
W Hz$^{-1}$ at $\sim$1\asec\ resolution, they found that {\it none} of the 
galaxies contain radio cores.  They detected nuclear emission in three galaxies,
but in all cases the morphology was diffuse, consistent with that seen in 
nearby circumnuclear starbursts such as NGC 253.  The VLA study of 
Filho, Barthel, \& Ho (2000) also failed to detect radio cores in a more
heterogeneous sample of 12 \hii\ nuclei. 

Information on nuclear X-ray cores in late-type galaxies is much more limited 
because to date there has been no systematic investigation of these systems 
with {\it Chandra}.  A few studies, however, have exploited the High 
Resolution Imager (HRI) on {\it ROSAT} to resolve the soft X-ray (0.5--2 keV) 
emission in late-type galaxies (Colbert \& Mushotzky 1999; Lira, Lawrence, \& 
Johnson 2000; Roberts \& Warwick 2000).  Although the resolution of the HRI 
($\sim$5\asec) is not ideal, it is nonetheless quite effective for identifying
point sources given the relatively diffuse morphologies of late-type galaxies.
Compact X-ray sources, often quite luminous (\gax 10$^{38}$ \lum), are 
frequently found, but generally they do {\it not} coincide with the 
galaxy nucleus; the nature of these ``ultraluminous X-ray sources'' 
is discussed by van~der~Marel (2004).

To summarize: unless \hii\ nuclei in late-type galaxies contain radio and 
X-ray cores far weaker than the current survey limits---a possibility worth 
exploring---they do not appear to conceal a significant population of 
undetected AGNs.  

\subsection{The $z \approx 0$ AGN Luminosity Function}

Many astrophysical applications of AGN demographics benefit from knowing
the AGN luminosity function, $\Phi(L,z)$\footnote{For the purposes of this 
paper, I will only consider the optical luminosity function.}.  Whereas 
$\Phi(L,z)$ has been reasonably well charted for high $L$ and high $z$ 
using quasars (Osmer 2004), it is very poorly known at low $L$ and low $z$.  
Indeed, until very recently there has been no reliable determination of 
$\Phi(L,0)$.

The difficulty in determining $\Phi(L,0)$ can be ascribed to a number of 
factors, as discussed in Huchra \& Burg (1992).  First and foremost is the 
challenge of securing a reliable, spectroscopically selected sample, as 
discussed in \S\S~1.2--1.4. Since nearby AGNs are expected to be faint relative 
to their host galaxies, most of the traditional techniques used to identify 
quasars cannot be applied without introducing large biases.  The faintness of 
nearby AGNs presents another obstacle, namely how to disentangle the nuclear 
emission---the only component relevant to the AGN---from the usually much 
brighter contribution from the host galaxy.  Finally, most optical luminosity 
functions of bright, more distant AGNs are specified in terms of the 
nonstellar optical continuum (usually the $B$ band), whereas spectroscopic 
surveys of nearby galaxies generally only reliably measure optical line 
emission (e.g., H\al) because the featureless nuclear continuum is often 
impossible to detect in ground-based, seeing-limited apertures.  

Huchra \& Burg (1992; see also Osterbrock \& Martel 1993) presented the first 
optical luminosity function of nearby Seyfert galaxies, based on the sample of 
AGNs selected from the CfA redshift survey.  They also calculated the 
luminosity function of LINERs, but it was known to be highly incomplete.  
Huchra \& Burg, however, did not have access to true nuclear luminosities for 
their sample; their luminosity function was based on {\it total}\ (nucleus 
plus host galaxy) magnitudes.

A different strategy can be explored by taking advantage of the fact that 
H\al\ luminosities are now available for nearly all of the AGNs in the Palomar 
survey (Ho et al. 1997a, 2003).  Ho et al. (2004) begin by calculating the 
nuclear H\al\ luminosity function using Schmidt's (1968) $V/V_{\rm max}$ 
method, where for each source $V$ is the volume it occupies given its 
distance and $V_{\rm max}$ is the maximum volume it could occupy were it to 
lie within the flux limit of the survey, taken to be the larger of the two 
volumes as calculated from the total optical magnitude limit of the survey 
($B_T = 12.5$ mag) and the flux limit for detecting emission lines. Next, 
Ho et al. (2004) exploit the fact that luminous AGNs obey a tight correlation 
between the luminosity of the optical, nonstellar continuum and the luminosity 
of the hydrogen Balmer lines, a relation that follows naturally from 
simple photoionization arguments (Searle \& Sargent 1968; Weedman 1976; Yee 
1980; Shuder 1981).  Using {\it nuclear}\ continuum magnitudes extracted from 
high-resolution {\it HST}\ images, Ho \& Peng (2001) showed that 
low-luminosity AGNs, too, obey the correlation established by the more 
luminous sources, albeit with somewhat greater scatter.  

Figure 1.8 presents the $B$-band nuclear luminosity function for the Palomar 
AGNs, computed by translating the extinction-corrected H\al\ luminosities to 
$B$-band absolute magnitudes with the aid of the empirical calibration between 
H\bet\ luminosity and $M_B$ of Ho \& Peng (2001) and an assumed H\al/H\bet\ 
ratio of 3.1.  Two versions are shown, each representing an extreme view of 
what kind of sources should be regarded as {\it bona fide}\ AGNs.  The open 
circles include only type~1 nuclei, sources in which broad H\al\ emission was 
detected and hence whose AGN status is incontrovertible.  This may be regarded 
as the most conservative assumption and a lower bound, since we know that 
genuine narrow-lined AGNs do exist (e.g., M104 or NGC 4261).  The solid circles 
lump together all sources classified as LINERs, transition objects, or 
Seyferts, both type~1 and type~2.  This represents the most optimistic view 
and an upper bound, since undoubtedly {\it some}\ narrow-lined sources must be 
stellar in origin but masquerading as AGNs.  The true space density of local 
AGNs most likely lies between these two possibilities.  In either case, the 
differential luminosity function is reasonably well approximated by a single 
power law from $M_B\,\approx$ --5 to --18 mag, roughly of the form 
$\Phi \propto L^{-1.2\pm0.2}$.  The slope may flatten for $M_B$ \gax\ --7 mag, 
but the luminosity function is highly uncertain at the faint end because of 
density fluctuations in our local volume.

For comparison, I have overlaid the luminosity function of $z$ \lax\ 0.3 
quasars and Seyfert 1 nuclei as determined by K\"{o}hler et al. (1997) from the
Hamburg/ESO UV-excess survey\footnote{Scaled to our adopted cosmological 
parameters of $H_0$ = 75 \kms\ Mpc$^{-1}$, $\Omega_{\rm m} = 0.3$, and 
$\Omega_{\Lambda} = 0.7$.}.  This sample extends the luminosity function from 
$M_B \,\approx$ --18 to --26 mag.  Although the two samples do not strictly 
overlap in luminosity, it is apparent the two samples roughly merge, and that 
the break in the combined luminosity function most likely falls near 
$M_B^* \approx -19$ mag, where the space density 
$\phi \approx 1\times 10^{-4}$ Mpc$^{-3}$ mag$^{-1}$.  I also plotted the 
quasar luminosity function obtained from the 2dF quasar redshift survey (2QZ),  
after evolving it to $z=0$ following the luminosity evolution prescription 
of Boyle et al. (2000).  The faint-end slope matches that of the local value 
quite well, but the break the 2QZ luminosity function drops much more sharply 
than the local sample.

\begin{figure*}[t]
\includegraphics[width=0.85\columnwidth,angle=270,clip]{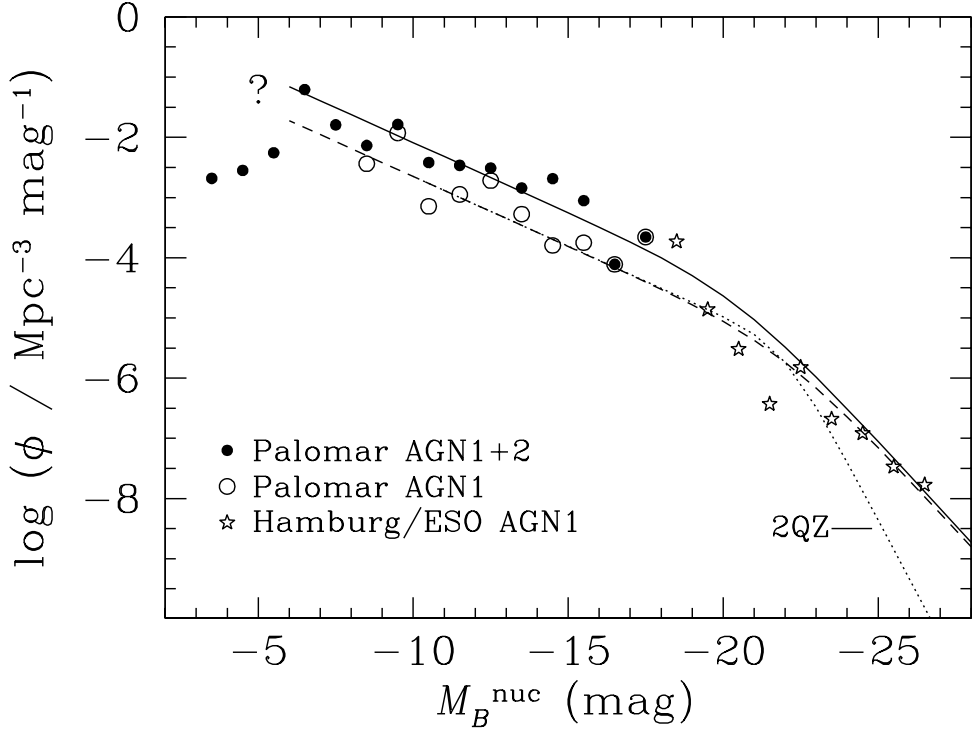}
\vskip 0pt \caption{
The $B$-band nuclear luminosity function of nearby AGNs derived 
from the Palomar survey.  The filled circles include all (type~1 + type~2) 
sources, while the open circles include only type~1 sources.  The 
sample of luminous Seyfert~1s and quasars from the Hamburg/ESO survey of 
K\"ohler et al. (1997) is shown as stars.  A double power-law fit to the 
Palomar and Hamburg/ESO samples is shown as a solid (type~1 + type~2) and 
dashed (type~1) curve.  The dotted curve represents the quasar luminosity 
function derived from the 2dF quasar redshift survey (2QZ), shifted to 
$z=0$ according to the luminosity evolution model of Boyle et al. (2000).
(Adapted from Ho et al. 2004.) 
\label{fig8}}
\end{figure*}

\subsection{Bolometric Luminosities and Eddington Ratios}

To gain further insight into the physical nature of nearby AGNs, it is more 
instructive to examine their bolometric luminosities, rather than their 
luminosities in a specific band or emission line.  Because AGNs emit a very 
broad spectrum, their bolometric luminosities ideally should be measured 
directly from their full spectral energy distributions (SEDs).  In practice, 
however, complete SEDs are not readily available for most AGNs, and one 
commonly estimates the bolometric luminosity by applying bolometric 
corrections derived from a set of well-observed calibrators.  As discussed in 
more detail in \S 1.5, the SEDs of low-luminosity AGNs differ quite markedly 
from those of conventionally studied AGNs.  Nonetheless, they do exhibit a 
characteristic shape, which enables bolometric corrections to be calculated.  
AGN researchers customarily choose as the reference point either the optical 
$B$ band or an X-ray band.  While the same strategy may be used for 
low-luminosity AGNs (e.g., Ho et al. 2000), it cannot yet be widely employed 
because nuclear optical or X-ray fluxes are not yet available for large 
samples.  What is available, by selection, is nuclear emission-line fluxes, 
and upper limits thereof.  Although the H\al\ luminosity comprises only a 
small percentage of the total power, its fractional contribution to the 
bolometric luminosity, as Ho (2003, 2004) notes, turns out to be fairly well 
defined.

Figure~1.9 shows the distributions of bolometric luminosities and their values
normalized with respect to the Eddington luminosity for Palomar galaxies with 
measurements of H\al\ luminosity and central stellar velocity dispersions. The
$M_\bullet - \sigma$ relation of Tremaine et al. (2002) was used to obtain 
\ledd.  Whereas LINER and transition nuclei both have a median \lbol\ 
$\approx$ 2\e{41} \lum, Seyfert nuclei are typically an order of magnitude 
more luminous (median \lbol\ $\approx$ 2\e{42} \lum).  The upper limits 
for the objects lacking any detectable line emission (absorption-line nuclei) 
cluster near \lbol\ $\approx$ 3\e{40} \lum.  These systematic trends persist 
when I consider the Eddington ratios.  One again, the distribution of 
\lbol/\ledd\ for LINERs is rather similar to that of transition objects 
(median \lbol/\ledd\ $\approx$ 2\e{-5} and 3\e{-5}, respectively), but both 
are quite distinct from Seyferts (median \lbol/\ledd\ $\approx$ 4\e{-4}).  
Notably, the vast majority of nearby nuclei have highly sub-Eddington 
luminosities.

\begin{figure*}[t]
\hspace*{-0.7cm}
\includegraphics[width=0.54\columnwidth,angle=270]{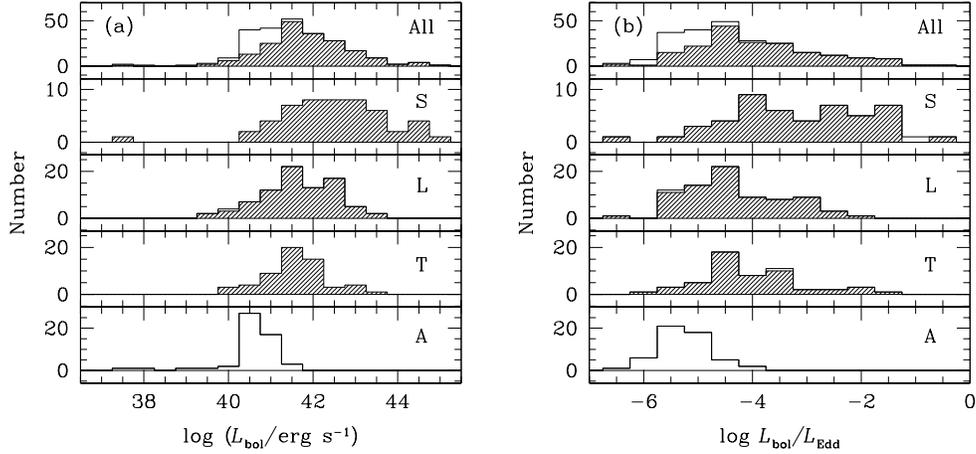}
\vskip 0pt \caption{
Distribution of ({\it a}) bolometric luminosity, $L_{\rm bol}$, and ({\it b})
ratio of bolometric luminosity to the Eddington luminosity, 
$L_{\rm bol}/L_{\rm Edd}$, for all objects, Seyferts (L), LINERs (L),
transition nuclei (T), and absorption-line nuclei (A).  The hatched and
open histograms denote detections and upper limits, respectively.
(Adapted from Ho 2004.) 
\label{fig9}}
\end{figure*}

\subsection{Host Galaxy Properties}

The near dichotomy in the distribution of Hubble types for galaxies hosting 
active versus inactive nuclei (Fig. 1.5) leads to the expectation that the two 
populations ought to have fairly distinctive global, and perhaps even nuclear, 
properties.  Moreover, a detailed examination of the host galaxies of AGNs 
may shed light on the origin of their spectral diversity.  These issues were
recently examined by Ho et al. (2003) using the database from the Palomar 
survey.  The main results are summarized here.

The host galaxies of Seyferts, LINERs, and transition objects display a 
remarkable degree of homogeneity in their large-scale properties.  After 
factoring out spurious differences arising from slight mismatches in Hubble 
type distribution, all three classes have essentially identical total 
luminosities ($\sim L^*$), bulge luminosities, sizes, and neutral gas 
content.  The only exception is that, relative to LINERs, transition objects 
may show a mild enhancement in the level of star formation, and they may be 
preferentially more inclined.  This is consistent with the hypothesis that the 
transition class arises from spatial blending of emission from a LINER and 
\hii\ regions.

Theoretical studies (e.g., Heller \& Shlosman 1994) suggest that large-scale
stellar bars can be highly effective in delivering gas to the central few
hundred parsecs of a spiral galaxy, thereby potentially leading to rapid star
formation.  Further instabilities result in additional inflow to smaller 
scales.  Thus, provided that an adequate reservoir of gas exists, the presence 
of a bar might be expected to influence the BH fueling rate, and hence the 
level of nonstellar activity.  The Palomar sample is ideally suited for 
statistical tests of this nature, which depend delicately on issues of sample 
selection effects and completeness.  Ho et al. (1997d) find that while the 
presence of a bar indeed does enhance both the probability and rate of star 
formation in galaxy nuclei, it appears to have no impact on either the 
frequency or strength of AGN activity.  Bearing in mind the substantial 
uncertainties introduced by sample selection (see discussion in Appendix B of 
Ho \& Ulvestad 2001), other studies broadly come to a similar conclusion (see 
review by Combes 2003).

In the same vein, dynamical interactions with neighboring companions should 
lead to gas dissipation, enhanced nuclear star formation, and perhaps central
fueling (e.g., Hernquist 1989).  Schmitt (2001) and Ho et al. (2003) studied 
this issue using the Palomar data, parameterizing the nearby environment of 
each object by its local galaxy density and the distance to its nearest 
sizable neighbor.  After accounting for the well-known morphology-density 
relation, it was found that the local environment, like bars, has little 
impact on AGNs.

The uniformity in host galaxy properties extends even to small, nuclear ($\sim
200$ pc) scales, in two important respects.  First, the velocity field of the
ionized gas, as measured by the width and asymmetry of the narrow emission
lines, appears to be crudely similar among the three classes, an observation
that argues against the proposition that fast shocks primarily drive the 
spectral variations observed in nearby galactic nuclei.  Second, the 
homogeneity among the three AGN classes is seen in their nuclear stellar 
content, which nearly always appears evolved.  The general dearth of young or 
intermediate-age stars presents a serious challenge to proposals that seek to 
account for the excitation of the emission lines in terms of starburst or 
post-starburst models.

\begin{figure*}[t]
\hspace*{+0.1cm}
\includegraphics[width=0.465\columnwidth,angle=0,clip]{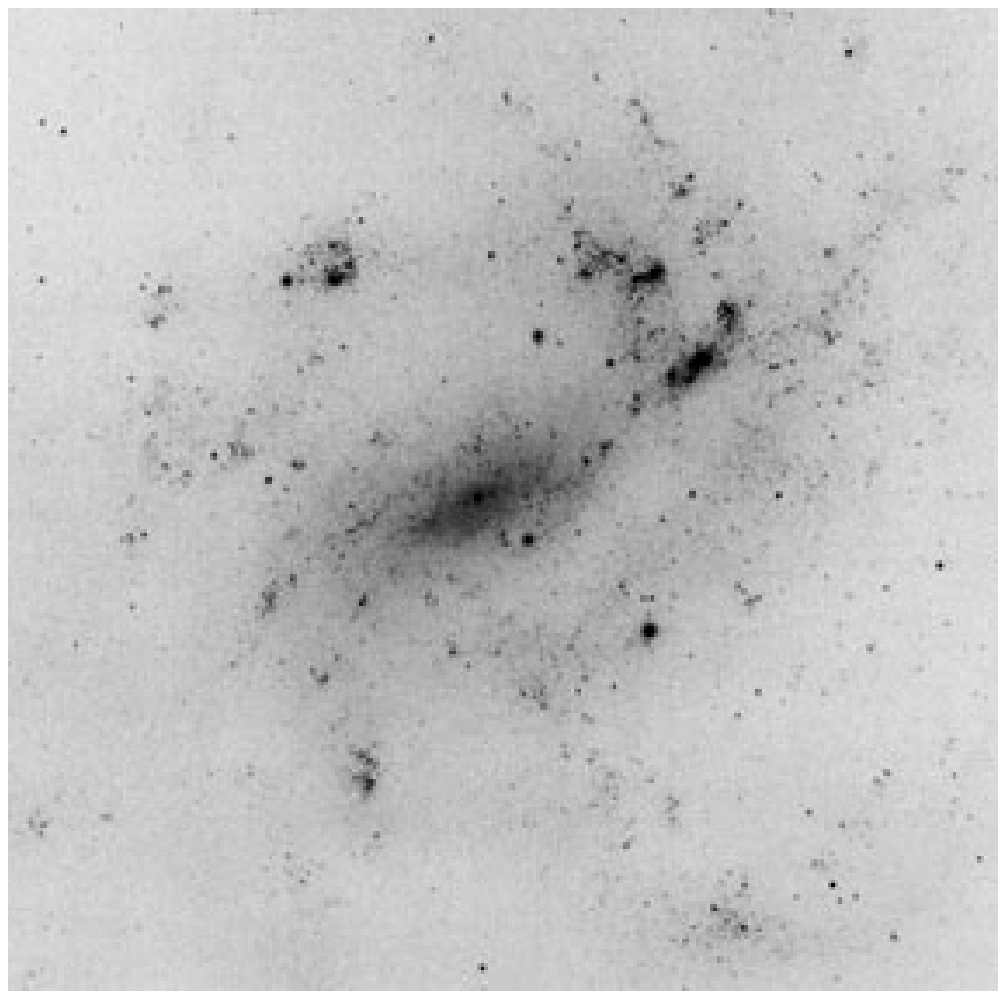}
\includegraphics[width=0.49\columnwidth,angle=0,clip]{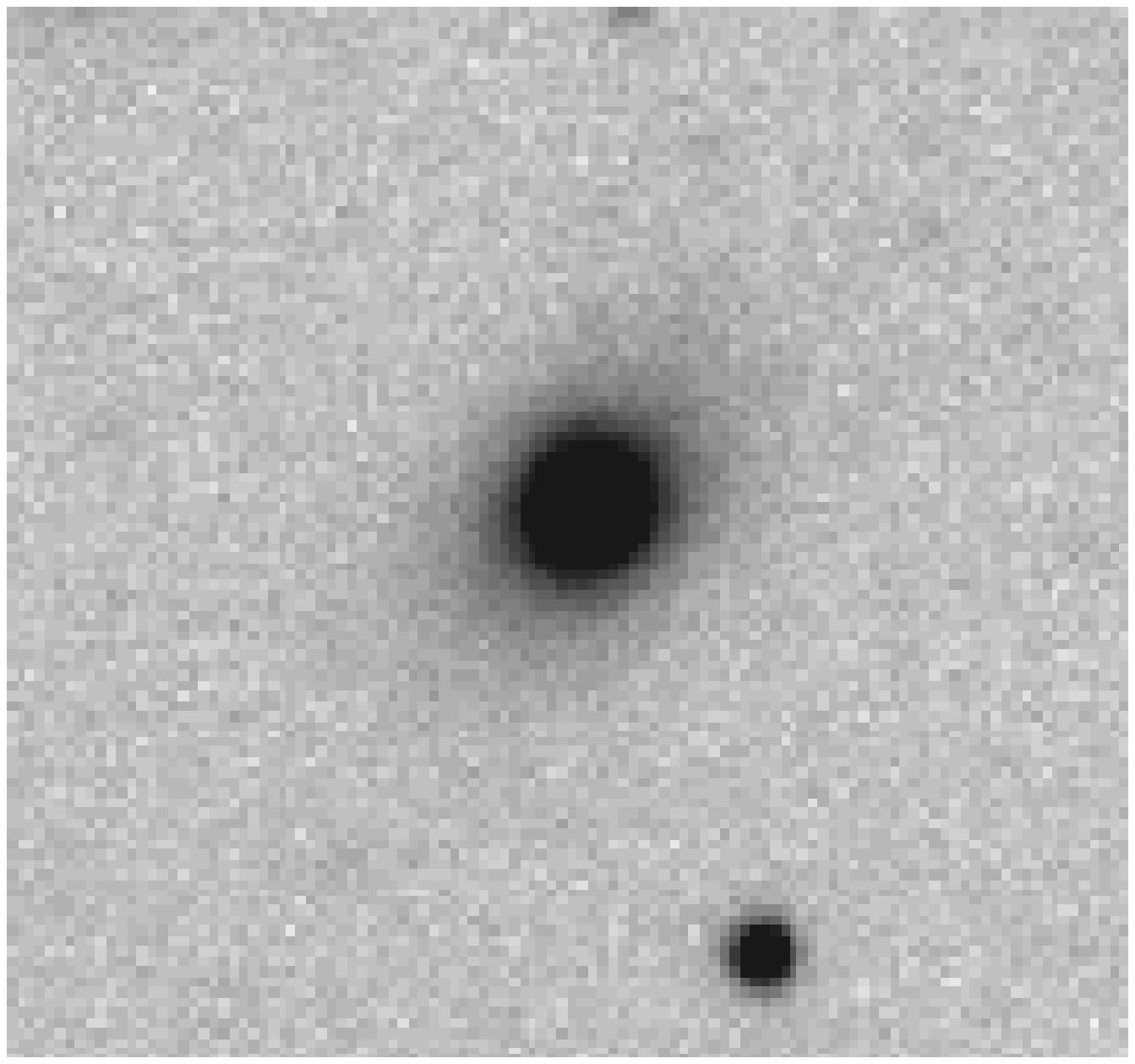}
\vskip 0pt \caption{
Two examples of AGNs in late-type galaxies.  The {\it left}\ panel shows an
optical image of NGC 4395, adapted from the Carnegie Atlas of Galaxies
(Sandage \& Bedke 1994); the image is $\sim$15\amin\ (17 kpc) on a side.
The {\it right}\ panel shows an $R$-band image of POX 52, adapted from Barth 
et al. (2004); the image is $\sim$25\asec\ (11 kpc) on a side.
\label{fig10}}
\end{figure*}

\subsection{Do Intermediate-mass Black Holes Exist?}

As summarized by Barth (2004) and Kormendy (2004), the observational evidence 
for supermassive BHs in the mass range \mbh\ $\approx\,10^6-10^{9.5}$ \solmass\ 
has become quite secure, to the point that important inferences on their 
demographics can be drawn (Richstone 2004).  Do central (nonstellar) BHs with 
masses below $10^6$ \solmass\ exist?  The current low end of the mass scale 
may reflect observational limitations rather than a true physical limit.  
There is certainly no compelling theory that prohibits the existence of BHs in 
the large gap between the stellar regime of 10 \solmass\ and $10^6$ \solmass.
The recent dynamical studies of Gerssen et al. (2002) and Gebhardt, Rich, \& 
Ho (2002) suggest that some massive star clusters may contain BHs in the mass 
range $10^3 < M_\bullet < 10^4$ \solmass\ (see review by van~der~Marel 2004).  

The existence of intermediate-mass BHs is further supported by the detection 
of AGNs in at least some very late-type and dwarf galaxies.  Two examples are 
particularly noteworthy (Fig. 1.10).  The nearby ($\sim$4 Mpc) galaxy NGC 4395 
contains all the usual attributes of a respectable AGN: broad optical and UV 
emission lines (Filippenko \& Sargent 1989; Filippenko, Ho, \& Sargent 1993), a 
compact radio core (Ho \& Ulvestad 2001; Wrobel, Fassnacht, \& Ho 2001), 
and rapidly variable hard X-ray emission (Moran et al. 1999, 2004; Shih, 
Iwasawa, \& Fabian 2003).  Contrary to expectations, however, NGC 4395 is a
bulgeless, extremely late-type (Sdm) spiral, whose central stellar velocity 
dispersion does not exceed $\sim 30$ \kms\ (Filippenko \& Ho 2003).  If 
NGC 4395 obeys the $M_\bullet - \sigma$ relation, its central BH should have a 
mass \lax $10^5$ \solmass.  This limit agrees surprisingly well with the value 
of \mbh\ estimated from its broad H\bet\ line width and X-ray variability 
properties ($\sim 10^4-10^5$ \solmass; Filippenko \& Ho 2003).

As first noted by Kunth, Sargent, \& Bothun (1987), the presence of a 
Seyfert-like nucleus in POX 52 is unusual because of the low luminosity 
of the host galaxy.  Barth et al. (2004) show that POX 52 bears a close 
spectroscopic resemblance to NGC 4395.  Based on the broad profile of H\bet, 
these authors derive a virial BH mass of 1.6\e{5} \solmass\ for POX 52, again 
remarkably close to the value of 1.3\e{5} \solmass\ predicted from the  
$M_\bullet - \sigma$ relation given the measured central stellar velocity 
dispersion of 35 \kms.  Deep images reveal POX 52 to be most akin to a dwarf 
elliptical galaxy, to date an unprecedented morphology for an AGN host galaxy.

The two objects highlighted above demonstrate that the mass spectrum of 
nuclear BHs very likely extends below $10^6$ \solmass.  Furthermore, they 
provide great leverage for anchoring the $M_\bullet - \sigma$ at the low end.  
But how common are such objects?  AGNs hosted in high-surface brightness, 
late-type spirals appear to be quite rare in the nearby Universe.  Within the 
comprehensive Palomar survey, NGC 4395 emerges as a unique case, and as argued 
in \S 1.4, the late-type galaxy population probably does not conceal a large 
number of misidentified AGNs.  The majority of late-type spirals do possess 
compact, photometrically distinct nuclei (B\"oker et al. 2002), 
morphologically not dissimilar from NGC 4395, but these nuclei are compact 
star clusters, not AGNs (Walcher et al. 2004).  The true incidence of AGNs 
like POX 52 is more difficult to assess.  Most spectroscopic surveys of 
blue-selected emission-line objects do not have sufficient signal-to-noise 
ratio to identify weak, broad emission lines.  On the other hand, Greene \& Ho 
(2004), in a preliminary analysis of the first data release from SDSS, have 
uncovered a number of broad-line AGNs in low-luminosity, presumably low-mass, 
host galaxies.  These appear to be excellent candidates for late-type galaxies 
with intermediate-mass BHs.

\section{The Nature of LINERs}

The recognition and definition of LINERs is based on their spectroscopic 
properties at optical wavelengths.  In addition to the AGN scenario, the 
optical spectra of LINERs unfortunately can be interpreted in several other 
ways that do not require an exotic energy source.  The principal alternatives
are shocks and hot stars (see reviews by Barth 2002 and Filippenko 2003).  As 
a consequence, it has 
often been suggested that LINERs may be a mixed-bag, heterogeneous collection 
of objects.  While the nonstellar nature of some well-studied LINERs is 
incontrovertible (e.g., M81, M87), the AGN content in the majority of LINERs 
continues to be debated.  Determining the physical origin of LINERs is more 
than of mere phenomenological interest.  Because LINERs are so numerous, they 
have repercussions on many issues related to AGN demographics.

\subsection{Evidence for an AGN Origin}

A number of recent developments provide considerable new insight into the
origin of LINERs.  I outline these below, and I use them to advance the
proposition that {\it most}\ LINERs are truly AGNs\footnote{Note that this
paper is concerned only with compact, {\it nuclear}\ LINERs ($r$ \lax 100 pc),
which are most relevant to the AGN issue.  I do not consider LINER-like
extended nebulae such as those associated with cooling flows, nuclear
outflows, starburst-driven winds, and circumnuclear disks.  LINERs selected
from samples of distant, interacting, or infrared-bright galaxies are
particularly vulnerable to confusion from these extended sources.  (See also 
discussion in \S~1.4.1.)}.  The discussion first focuses on ``pure'' LINERs, 
returning later to the class of transition objects, whose nature remains more 
obscure.

\begin{enumerate}

\item
{\it{Detection of broad emission lines:}}
Luminous, unobscured AGNs distinguish themselves unambiguously by their 
characteristic broad permitted lines.  The detection of broad H\al\ emission 
in $\sim$25\% of LINERs (Ho et al. 1997e) thus constitutes strong evidence in 
favor of the AGN interpretation of these sources.  LINERs, like Seyferts, 
evidently come in two flavors---some have a visible BLR (type~1), and others 
do not (type~2).  The broad component becomes progressively more difficult to
detect in ground-based spectra for permitted lines weaker than H\al.  Where 
available, however, broad, higher-order Balmer lines as well as UV lines such 
as Ly$\alpha$, \civ\ \lamb1549, \mgii\ \lamb2800, and \feii\ multiplets can be 
seen in \hst\ spectra (Barth \etal 1996; Ho \etal 1996).  I further recall
the double-peaked broad lines mentioned in \S~1.5.3.

\item
{\it{Detection of hidden BLRs:}}
An outstanding question, however, is what fraction of the more numerous 
LINER~2s are AGNs.  By analogy with the Seyfert~2 class, surely {\it some}\ 
LINER~2s must be genuine AGNs---that is, LINERs whose BLR is obscured along 
the line of sight to the viewer.  There is no {\it a priori}\ reason why the 
unification model, which has enjoyed such success in the context of 
Seyfert galaxies, should not apply equally to LINERs.  The existence of an 
obscuring torus does not obviously depend on the ionization level of the NLR.  
If we suppose that the ratio of LINER~2s to LINER~1s is the same as the ratio 
of Seyfert~2s to Seyfert~1s, that ratio being 1.6:1 in the Palomar survey, we 
can reasonably surmise that the AGN fraction in LINERs may be as high as 
$\sim$60\%.  That at least some LINER~2s {\it do}\ contain a hidden BLR was 
demonstrated by the spectropolarimetric observations of Barth, Filippenko, \& 
Moran (1999a, b).

\item
{\it{Naked LINER~2s:}}
The BLR in some LINER~2s may be intrinsically absent, not obscured.  If BLR 
clouds arise from condensations in a radiation-driven, outflowing wind (e.g., 
Murray \& Chiang 1995), a viewpoint now much espoused, then it is reasonable 
to expect that very low-luminosity sources would be incapable of generating a 
wind, and hence of sustaining a BLR.  A good example of such a case is 
NGC 4594 (the ``Sombrero'' galaxy).   Its nucleus, although clearly an AGN, 
shows no trace of a broad-line component, neither in direct light (Ho et al. 
1997e), not even when very well isolated with a small \hst\ aperture 
(Nicholson et al. 1998), nor in polarized light (Barth et al. 1999b).  Its 
Balmer decrement indicates little reddening to the NLR.  For all practical 
purposes, the continuum emission from the nucleus looks unobscured: it is 
detected in the UV (Maoz et al. 1998) and in the soft and hard X-rays (Fabbiano
\& Juda 1997; Nicholson et al. 1998; Ho et al. 2001; Pellegrini et al. 2002; 
Terashima et al. 2002), with evidence for only moderate intrinsic absorption 
(Pellegrini et al. 2003).  In short, there is no sign of anything being hidden 
or much doing the hiding.  So where is the BLR?  It is not there.  A number of 
authors have also emphasized the existence of unabsorbed Seyfert~ 2 nuclei 
(e.g., Panessa \& Bassani 2002; Gliozzi et al. 2004).  These considerations 
lead to the conclusion that the mere absence of a BLR does not constitute 
evidence against the AGN pedigree of an object.

\item
{\it{Compact cores:}}
AGNs, at least when unobscured, traditionally reveal themselves as pointlike 
nuclear sources at virtually all wavelengths.  This fact can be exploited 
by searching for compact cores in LINERs.  To overcome the contrast problem, 
imaging observations of this kind are only meaningful for sufficiently high 
angular resolution.  Moreover, in practice certain spectral windows are 
more advantageous than others.  Nuclear point sources turn out to be 
surprisingly difficult to extract from optical and near-infrared images of 
nearby galaxies, even at the 0\farcs1 resolution of \hst\ (see, e.g., Ho \& 
Peng 2001; Ravindranath et al. 2001; Peng et al. 2002).  This is due to a
number of factors, chiefly the dominance of the host galaxy bulge and the 
complexity of dust structures in nuclear regions at these wavelengths.  
The bulge light largely disappears in the UV, but the detection rate of 
nuclear cores in the UV is only $\sim$25\% for LINERs (Maoz et al. 1995; 
Barth et al. 1998).  This band is especially hard to work with because 
it is particularly susceptible to dust extinction, to confusion from 
young stars when present (Maoz et al. 1998), and to intrinsic variations due 
to the form of the SEDs of low-luminosity AGNs (Ho 1999b; see point 5 below).

Compact nuclei can be detected most cleanly in the X-rays and radio (see also 
\S~1.4.4).  These regions are least sensitive to obscuration and provide the 
highest contrast between nonstellar and stellar emission.  There have been a 
number of attempts to systematically investigate LINERs using X-ray data 
obtained from {\it ROSAT}\ (Koratkar et al. 1995; Komossa, B\"ohringer, \& 
Huchra 1999; Roberts \& Warwick 2000; Halderson et al. 2001; Roberts, Schurch, 
\& Warwick 2001), {\it ASCA}\ (Ptak et al.  1999; Terashima, Ho, \& Ptak 2000; 
Terashima et al. 2000, 2002), and {\it BeppoSAX}\ (Georgantopoulos et al. 
2002). While these efforts have been enormously useful in delineating the basic 
X-ray properties of LINERs, particularly in the spectral domain, they suffer 
from two crucial limitations.  First, the angular resolutions of all the above 
X-ray facilities, with the possible exception of the {\it ROSAT}/HRI under 
some circumstances (\S 1.4.4), are grossly incapable of properly isolating any 
but the brightest nuclei. And second, the faintness of the typical targets 
compels most investigators to study only limited, inevitably biased, samples. 

The advent of {\it Chandra}\ has dramatically improved this situation.  The 
high angular resolution of the telescope ($\sim$1\asec) and the low background 
noise of the CCD detectors allow faint point sources to be detected with brief 
(few ks) exposures (Ho et al. 2001; Terashima \& Wilson 2003).  This makes 
feasible, for the first time, X-ray surveys of large samples of galaxies 
selected at non-X-ray wavelengths. Ho et al. (2001) used the ACIS camera to 
image a distance-limited sample of Palomar AGNs.  Their analysis of a 
preliminary subset indicates that X-ray cores, some as faint as $\sim 10^{38}$ 
\lum\ in the 2--10 keV band, are found in $\sim$75\% of LINERs; the detection 
rate is roughly similar for LINER~1s and 2s.

As in the X-rays, AGNs nearly universally emit at some level in the radio.  
Although most AGNs are radio quiet, they are seldom radio silent when observed 
with sufficient sensitivity and angular resolution.  For example, Seyfert 
galaxies generally contain radio cores, which are often accompanied by linear, 
jetlike features (e.g., Ulvestad \& Wilson 1989; Kukula et al. 1995; Thean et 
al. 2000; Ho \& Ulvestad 2001; Schmitt et al. 2001).  The complete sample of 
Seyferts selected from the Palomar survey shows a detection rate of $\sim$80\% 
at 5~GHz and 1\asec\ resolution (Ho \& Ulvestad 2001; Ulvestad \& Ho 2001a); 
their radio powers span $10^{18}-10^{21}$ W Hz$^{-1}$.  

LINERs have been surveyed somewhat less extensively than Seyferts.  As 
mentioned in \S~1.4.2, radio interferometric surveys of nearby elliptical and 
S0 galaxies detect a high fraction of low-power cores, most of which are 
optically classified as LINERs.  VLA studies of well-defined subsamples of 
LINERs chosen from the Palomar survey show qualitatively similar trends 
(Van~Dyk \& Ho 1997; Nagar et al. 2000, 2002).  At 5 and 8 GHz, where the 
sensitivity is highest, $\sim$60\%--80\% of LINERs, independent of type, 
contain radio cores.  VLBI observations of the brighter, VLA-detected sources 
generally reveal brightness temperatures \gax $10^{6-8}$ K (Falcke et al. 
2000; Ulvestad \& Ho 2001b; Filho, Barthel, \& Ho 2002b; Anderson, Ulvestad, 
\& Ho 2004).

To summarize: the majority of LINERs, both type~1 and 2, photometrically 
resemble AGNs insofar as they emit compact, pointlike hard X-ray and radio 
emission.  

\item
{\it{Spectral energy distributions:}}
The broad-band spectrum of luminous, unabsorbed AGNs follows a fairly 
universal shape (e.g., Elvis \etal 1994).  The SED from the infrared to 
the X-rays can be roughly represented as the sum of an underlying power law 
($L_{\nu}\,\propto\,\nu^{-1}$) and a few distinct components, the most 
prominent of which is the ``big blue bump'' usually attributed to thermal 
emission from an optically thick, geometrically thin accretion disk (Shields 
1978; Malkan \& Sargent 1982).  The SEDs of LINERs deviate markedly from the 
standard form of high-luminosity AGNs (Ho 1999b, 2002b; Ho et al. 2000), as 
shown in Figure 1.11.   The most conspicuous difference can be seen in the 
apparent absence of a UV excess.  The SEDs of these sources also tend to be 
generically ``radio loud,'' defined here by the convention that the 
radio-to-optical luminosity ratio exceeds a value of 10.  In fact, radio 
loudness seems to be a property common to essentially {\it all}\ nearby weakly 
active nuclei (Ho 2002a) and a substantial fraction of Seyfert nuclei (Ho \& 
Peng 2001).   Using a definition of radio loudness based on the relative 
strength of the X-ray and radio emission, Terashima \& Wilson (2003) also 
find that LINERs tend to be radio loud.  

While the SEDs of LINERs differ from those of those of traditional AGNs, it is 
important to emphasize that they {\it do}\ approximate the SEDs  predicted for 
radiatively inefficient accretion flows onto BHs (e.g., Quataert et al.  1999; 
Ptak et al. 2004).  At the same time, they definitely bear little resemblance 
to SEDs characteristic of ``normal'' stellar systems (see, e.g., Schmitt et 
al. 1997).  Inactive galaxies or starburst systems not strongly affected by 
dust extinction emit the bulk of their radiation in the optical--UV and in the 
thermal infrared regions, with only an energetically miniscule contribution 
from X-rays (Fig. 1.11).   Indeed, the decidedly {\it nonstellar} nature of the 
SEDs of LINERs can be regarded as compelling evidence that LINERs are 
accretion-powered sources, albeit of an unusual ilk.

\begin{figure*}[t]
\includegraphics[width=0.73\columnwidth,angle=-90]{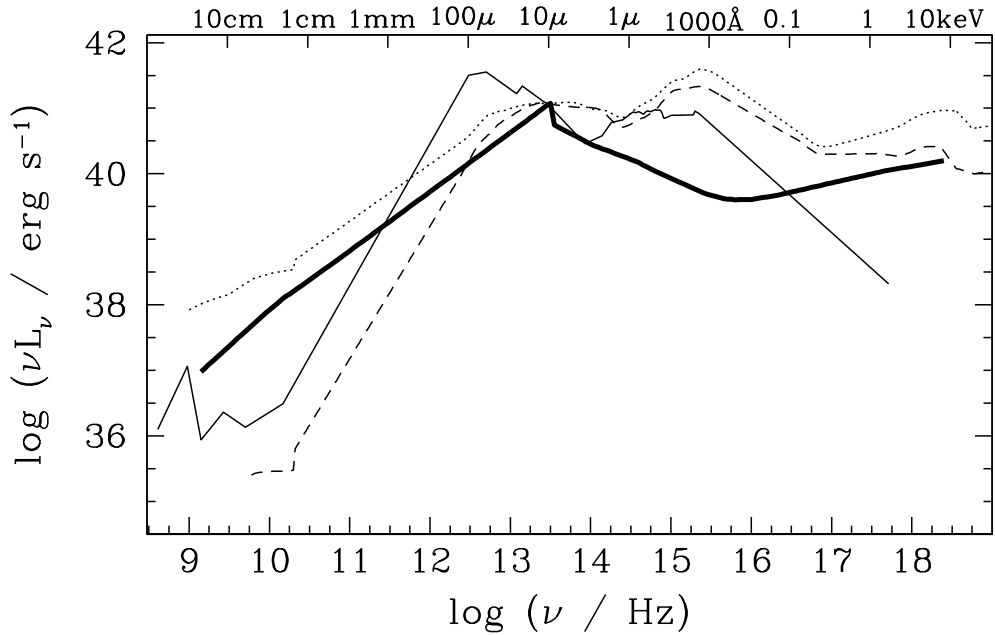}
\vskip 0pt \caption{
The average SED of low-luminosity AGNs ({\it heavy solid line}), adapted from 
Ho (1999b).  Overplotted for comparison are the average SEDs of powerful 
radio-loud ({\it dotted line}) and radio-quiet ({\it dashed line}) AGNs (Elvis 
\etal 1994), and of low-extinction starburst galaxies ({\it light solid line}; 
Schmitt et al. 1997).  The curves have been arbitrarily normalized to 
the luminosity at 10 $\mu$m.
\label{fig11}}
\end{figure*}

\item
{\it{Host galaxies:}}
As discussed in \S~1.4.7,  LINERs and Seyferts live in virtually identical 
host galaxies.  To the extent that Seyferts are regarded as AGNs, the 
close similarity of their hosts with those of LINERs lends supporting, 
if albeit indirect, evidence that the two classes share a common origin.

\item
{\it{Detection of massive BHs:}}
Finally, I note the obvious fact that a significant fraction of the galaxies
with detected BHs are, in fact, well-known LINERs. These include M81, M84, M87, 
NGC 4261, and NGC 4594.  Although certainly no statistical conclusions can yet 
be drawn from such meager statistics, these examples nevertheless illustrate 
that at least some LINERs seem to be directly connected with BH accretion.

\end{enumerate}

\subsection{Excitation Mechanisms}

The above arguments lend credence to the hypothesis that a sizable fraction of 
LINERs---indeed the {\it majority}---are directly related to AGNs.  In this 
context, photoionization by a central AGN surfaces as the most natural 
candidate for the primary excitation mechanism of LINERs. The optical spectra
of LINERs can be readily reproduced in AGN photoionization calculations by 
adjusting the ionization parameter, $U$, defined as the ratio of the density 
of ionizing photons to the density of nucleons at the illuminated face of a 
cloud.  Whereas the NLR spectrum of Seyferts can be well fitted with 
$\log U \approx -2.5\pm0.5$ (e.g., Ferland \& Netzer 1983; Stasi\'nska 1984; 
Ho, Shields, \& Filippenko 1993b), that of LINERs requires $\log U \approx 
-3.5\pm1.0$ (Ferland \& Netzer 1983; Halpern \& Steiner 1983; P\'equignot 
1984; Binette 1985; Ho et al.  1993a).  What factors contribute to the lower 
ionization parameters in LINERs?  Ho et al. (2003) identify the central 
luminosity and gas density as two relevant factors.  According to the 
statistics from the Palomar survey, LINERs on average have lower luminosities 
and lower gas densities than Seyferts.  For a given volume filling factor, 
this leads to lower ionization parameters in LINERs compared to Seyferts, 
although not at a level sufficient to account for the full difference between 
the two classes.

Despite the natural appeal of AGN photoionization, alternative excitation 
mechanisms for LINERs have been advanced.  Collisional ionization by shocks 
has been a popular contender from the outset (Koski \& Osterbrock 1976; 
Fosbury et al. 1978; Heckman 1980b; Dopita \& Sutherland 1995; Alonso-Herrero 
et al. 2000; Sugai \& Malkan 2000).  Dopita \& Sutherland (1995) showed that 
the diffuse radiation field generated by fast ($\upsilon\,\approx$ 150--500 
\kms) shocks can reproduce the optical narrow emission lines seen in both 
LINERs and Seyferts.  In their models, LINER-like spectra are realized under 
conditions in which the precursor \hii\ region of the shock is absent, as 
might be the case in gas-poor environments.  The postshock cooling zone 
attains a much higher equilibrium electron temperature than a photoionized 
plasma; consequently, a robust prediction of the shock model is that shocked 
gas should produce a higher excitation spectrum, most readily discernible in 
the UV, than photoionized gas.  In all the cases studied so far, however, the 
UV spectra are inconsistent with the fast-shock scenario because the observed 
intensities of the high-excitation lines such as \civ\ \lamb1549 and \heii\ 
\lamb1640 are much weaker than predicted (Barth \etal 1996, 1997; Maoz \etal 
1998; Nicholson \etal 1998; Gabel et al. 2000).  Dopita \etal (1997) used the 
spectrum of the circumnuclear {\it disk} of M87 to advance the view that 
LINERs are shock excited.  This argument is misleading because their analysis 
deliberately avoids the nucleus.  Sabra et al. (2003) demonstrate that the 
UV--optical spectrum of the {\it nucleus}\ of M87 is best explained by a 
multi-component photoionization model.

A recent analysis of the emission-line profiles of the Palomar nuclei further 
casts doubt on the viability of the shock scenario (Ho et al. 2003).
The velocity dispersions of the nuclear gas generally fall short of the values 
required for shock excitation to be important.  Furthermore, the close 
similarity between the velocity field of LINERs and Seyferts, as deduced from 
their line profiles, contradicts the basic premise that shocks are primarily 
responsible for the spectral differences between the two classes of objects.

Another widely discussed class of models invokes hot stars, formed in a 
short-duration burst of star formation, to supply the primary ionizing 
photons.  Ordinary O-type stars with effective temperatures typical of those 
inferred in giant \hii\ regions in galactic disks do not produce sufficiently 
strong low-ionization lines to account for the spectra of LINERs.  The 
physical conditions in galactic nuclei, on the other hand, may be more 
favorable for generating LINER-like spectra.  For example, Terlevich \& 
Melnick (1985) postulate that the high-metallicity environment of galactic 
nuclei may be particularly conducive to forming very hot [$T \approx (1-2) 
\times 10^5$ K], luminous Wolf-Rayet stars, whose ionizing spectrum would 
effectively mimic the power-law continuum of an AGN. The models of Filippenko 
\& Terlevich (1992) and Shields (1992) appeal to less extreme conditions.  
These authors show that photoionization by ordinary O stars embedded in 
an environment with high density and low ionization parameter can 
explain the spectral properties of transition objects.  Barth \& Shields 
(2000) extended this work by modeling the ionizing source not as single 
O-type stars but as a more realistic evolving young star cluster.  They 
confirm that young, massive stars can indeed generate optical emission-line 
spectra that match those of transition objects, and, under some plausible 
conditions, even those of {\it bona fide} LINERs.  But there is an 
important caveat: the star cluster must be formed in an instantaneous burst, 
and its age must coincide with the brief phase ($\sim$3--5 Myr after the 
burst) during which sufficient Wolf-Rayet stars are present to supply the 
extreme-UV photons necessary to boost the low-ionization lines.  The 
necessity of a sizable population of Wolf-Rayet stars is also emphasized 
in the recent study by Gabel \& Bruhweiler (2002).  As 
discussed in Ho et al. (2003), the main difficulty with this scenario, and 
indeed with all models that appeal to young stars (e.g., Taniguchi, Shioya, \& 
Murayama 2000), is that the nuclear stellar population of the host galaxies 
of the majority of nearby AGNs, irrespective of spectral class, is 
demonstrably {\it old}.  Stellar absorption indices indicative of young 
or intermediate-age stars are seldom seen, and the telltale emission features
of Wolf-Rayet stars are notably absent.  These empirical facts seriously 
undermine stellar-based photoionization models of AGNs.

\subsection{Transition Objects}

The physical origin of transition nuclei continues to be a thorny, largely 
unresolved problem. Since these objects make up a significant fraction of 
emission-line nuclei, this complication unfortunately casts some uncertainty 
into the demography of local AGNs.

In two-dimensional optical line-ratio diagrams (Fig. 1.3), transition nuclei 
are empirically defined to be those sources that lie sandwiched between the 
loci of ``normal'' \hii\ regions and LINERs.  This motivated Ho et al. (1993a) 
to proposed that transition objects may be composite systems consisting of a 
LINER nucleus plus an \hii\ region component.  The latter could arise from 
neighboring circumnuclear \hii\ regions or from \hii\ regions randomly 
projected along the line of sight.  A similar argument, based on decomposition 
of line profiles, has been made by V\'eron, Gon\c{c}alves, \& V\'eron-Cetty 
(1997) and Gon\c{c}alves, V\'eron-Cetty, \& V\'eron (1999).   

If transition objects truly are LINERs sprinkled with a frosting of star 
formation, one would expect that their host galaxies should be largely similar 
to those of LINERs, namely bulge-dominated systems (\S~1.4.7), modulo minor 
differences due to the ``excess'' contaminating star formation.  The study of 
Ho et al. (2003) largely supports this picture.  The host galaxies of 
transition nuclei exhibit systematically higher levels of recent star 
formation, as indicated by their far-infrared emission and broad-band optical 
colors, compared to LINERs of matched morphological types.  Moreover, the host 
galaxies of transition nuclei tend to be slightly more inclined than LINERs.  
Thus, all else being equal, transition-type spectra seem to be found precisely 
in those galaxies whose nuclei have a higher probability of being contaminated 
by extra-nuclear emission from star-forming regions.

This story, however, has some holes.  If simple spatial blending of 
circumnuclear \hii\ regions is sufficient to transform a regular LINER into a 
transition object, the LINER nucleus should reveal itself unambiguously in
spectra taken with angular resolution sufficiently high to isolate it.  
This test was performed by Barth, Ho, \& Filippenko (2003), who obtained 
{\it HST}/STIS spectra, taken with a 0\farcs2-wide slit, of a well-defined 
subsample of 15 transition objects selected from the Palomar catalog.  
To their surprise, the small-aperture spectra of the nuclei, for the most 
part, look very similar to the ground-based spectra; they are {\it not} more 
LINER-like.

The ``masqueraded-LINER'' hypothesis can be further tested by searching for 
compact radio and X-ray cores using high-resolution images.  Recall that this 
is a highly effective alternative method to filter out weak AGNs (\S~1.4.4 
and \S~1.5.1).  Filho et al. (2000, 2002a, 2004) have systematically surveyed 
the full sample of Palomar transition objects using the VLA.  They find that 
$\sim$25\% of the population contains arcsecond-scale radio cores.  These 
cores appear to be largely nonstellar in nature.  The brighter subset of 
these sources that are amenable to follow-up VLBI observations (Filho et al. 
2004) all reveal more compact (milliarcsecond-scale) cores with flat radio 
spectra and high brightness temperatures ($T_B$ \gax\ $10^7$ K).  In their 
preliminary analysis of a {\it Chandra} survey of Palomar galaxies, Ho et al. 
(2001) noted that transition objects show a marked deficit of X-ray cores.  
Although based on small-number statistics, the frequency of X-ray cores for 
the transition objects in the Ho et al. study also turns out to be 25\%.

The above considerations suggest a conservative lower limit of $\sim$25\% 
for the AGN fraction in transition objects, or a reduction from 13\% to 
$\sim$3\% of the overall galaxy population.  In turn, the total AGN fraction 
(LINERs, Seyferts, and accretion-powered transition objects) for all 
local galaxies decreases from 43\% to 33\%.  These revised rates are 
lower limits because of the imperfect correspondence between ``genuine'' AGNs 
and the presence of radio and X-ray cores.  After all, for reasons that are 
not yet understood, clearly not all Seyfert nuclei are detected in the radio 
(e.g., Ho \& Ulvestad 2001; Ulvestad \& Ho 2001), and X-rays in the 2--10 keV 
band will be extinguished for gas with sufficiently large column densities 
(\gax $10^{24}$ cm$^{-2}$).  Such low-luminosity Compton-thick nuclei, if 
present, can be uncovered with sensitive, high-resolution observations at 
harder X-ray energies.

If the majority of transition objects are not AGNs, we are faced with a new 
conundrum.  What are they?  For the reasons explained above, the source of 
their line excitation is unlikely to be shock heating or photoionization by 
hot, massive stars.  Here I suggest two possibilities worth 
considering.  First, the ionizing radiation field might originate from hot, 
{\it evolved} stars.  This idea has been advocated by Binette et al. (1994), 
who proposed that post-asymptotic giant branch stars, which can attain 
effective temperatures as high as $\sim 10^5$ K, might be responsible for 
photoionizing the extended ionized gas often observed in elliptical galaxies.  
The emission-line spectrum of these nebulae, in fact, tend to be of relatively 
low ionization (Demoulin-Ulrich, Butcher, \& Boksenberg 1984; Phillips 
et al. 1986; Zeilinger et al. 1996).   Invoking evolved stars has the obvious 
appeal of not conflicting with the dominant old stellar population found in 
the centers of nearby galaxies.   Second, the integrated (off-nuclear) X-ray 
emission of the central regions of galaxies may contribute nonnegligibly to 
the ionizing photon budget.  Recent {\it Chandra}\ and {\it XMM-Newton}\ 
images of the centers of nearby, ``ordinary'' galaxies have resolved the X-ray 
emission into two components: discrete sources and diffuse, hot gas.  The 
discrete X-ray source population consists mainly of X-ray binaries, mostly of 
the low-mass variety (see Fabbiano \& White 2004 for a review).  While 
X-ray--emitting plasma has long been known to be pervasive in giant elliptical 
galaxies, it now appears that it may be a generic constituent even in 
spheroids of lower mass.  For example, diffuse, hot gas has been detected in 
the central regions of the Milky Way (Baganoff et al.  2004), M31 (Shirey et 
al. 2001), and M32 (Ho, Terashima, \& Ulvestad 2003).  Since X-ray binaries 
and X-ray--emitting gas have ``hard'' spectra (compared to, say, O-type stars), 
they would naturally be conducive to producing strong low-ionization optical 
lines when used as an ionizing source.  These unconventional sources of 
ionization---hot, evolved stars, X-ray binaries, and X-ray plasma---{\it must} 
contribute at some level, insofar as we know empirically that they exist.  
Their ubiquitous presence likely maintains a pervasive, diffuse, ionizing 
radiation field, which may be sufficient to sustain a ``baseline'' level of 
weak optical line emission.  It would be fruitful to further explore these 
issues quantitatively with photoionization models. 

\section{Summary}

This review argues that the demographics of AGN activity in nearby galaxies 
can inform us much about the demographics of massive BHs.  While ultimately 
there is no substitute for direct dynamical mass measurements, such an 
approach is often neither practical nor feasible.  AGN statistics provide 
important complementary information.  The following points are the most 
germane to BH demography.

\begin{enumerate}
 
\item
Nuclear activity is extremely common in the nearby Universe.  Over 40\% of 
all nearby galaxies qualify as AGNs or AGN candidates according to their 
emission-line spectral properties.  

\item
LINERs are the most common variety of local AGN candidates.

\item
The majority of LINERs appear to be genuinely accretion-powered systems.
Thus, most LINERs should be considered AGNs.

\item
Nuclear activity preferentially occurs in bulge-dominated galaxies.  Galaxies 
with Hubble types later than Sbc become progressively dominated by nuclear 
star formation.

\item
The physical origin of the so-called transition objects remains largely 
unknown, although at least 25\% of them appear to be AGNs.  This uncertainty 
affects the quantitative conclusions from (1) and (2), but the qualitative 
picture remains unchanged.

\item
Inasmuch as central BHs are a precondition for AGN activity, the detection 
rate of AGNs establishes a lower limit on the incidence of massive BHs in 
nearby galactic nuclei.  The above findings support the prevailing belief, 
based on dynamical studies and energy arguments (see Barth 2004; Kormendy 
2004; Richstone 2004), that massive BHs are a ubiquitous feature of massive 
galaxies.  

\item
Central massive BHs, while perhaps uncommon in pure-disk or dwarf galaxies, do 
not completely shun such environments.  Some late-type galaxies definitely 
harbor lightweight nuclear BHs, which extend the BH mass function down to 
the regime of \mbh\ $\approx\, 10^4-10^5$ \solmass, although the frequency of 
such objects is not yet well established.  These intermediate-mass 
BHs seem to obey the \mbh-$\sigma$ relation established by the 
supermassive ($10^6-10^{9.5}$ \solmass) BHs.

\item
Local AGNs are generically weak, characterized by highly sub-Eddington 
luminosities.  Most central BHs in nearby galaxies are either quiescent or 
only weakly active.

\end{enumerate}

\vspace{0.3cm}
{\bf Acknowledgements}.
My research is supported by the Carnegie Institution of Washington and by 
NASA grants from the Space Telescope Science Institute (operated by AURA, 
Inc., under NASA contract NAS5-26555).  I would like to recognize the 
significant contributions of my collaborators, especially A. J. Barth, A. V.
Filippenko, D. Maoz, E. C. Moran, C. Y. Peng, A. Ptak, H.-W.  Rix, W. L. W. 
Sargent, J. C. Shields, Y. Terashima, and J. S. Ulvestad.  I thank A. J. 
Barth, A. V. Filippenko, and W. L. W. Sargent for permission to cite material 
in advance of publication.  A. J. Barth, A. V. Filippenko, and J. S. Mulchaey 
gave thoughful comments on the manuscript.

\end{document}